\documentclass[fleqn]{article}
\usepackage{longtable}
\usepackage{graphicx}

\begin {document}
\begin{flushleft}
{\LARGE
{\bf Energy levels, radiative rates  and lifetimes for transitions in Br-like ions with 38 $\le$ Z $\le$ 42}
}\\

\vspace{1.5 cm}

{\bf {Kanti  M  ~Aggarwal and  Francis   P   ~Keenan}}\\ 

\vspace*{1.0cm}

Astrophysics Research Centre, School of Mathematics and Physics, Queen's University Belfast, Belfast BT7 1NN, Northern Ireland, UK\\ 
\vspace*{0.5 cm} 

e-mail: K.Aggarwal@qub.ac.uk \\

\vspace*{0.20cm}

Received  5 March 2014\\  Accepted for publication 26 September  2014 \\ Published xx Month  2014 \\ Online at stacks.iop.org/PhysScr/vol/number

\vspace*{1.5cm}

PACS Ref: 31.25 Jf, 32.70 Cs,  95.30 Ky

\vspace*{1.0 cm}

\hrule

\vspace{0.5 cm}

\end{flushleft}

\clearpage


\begin{abstract}

Energy levels and radiative rates for transitions in five Br-like ions (Sr IV, Y V, Zr VI, Nb VII and Mo VIII) are calculated with the general-purpose relativistic atomic structure package ({\sc grasp}). Extensive configuration interaction has been included and results are presented among the lowest 31 levels of the 4s$^2$4p$^5$, 4s$^2$4p$^4$4d and 4s4p$^6$    configurations. Lifetimes for these levels have also been determined, although unfortunately no measurements are available with which to compare. However, recently theoretical results have been reported by Singh {\em et al} [Phys. Scr. {\bf 88} (2013) 035301] using the same {\sc grasp} code. But their reported data for radiative rates and lifetimes cannot be reproduced and  show discrepancies of up to five orders of magnitude with the present calculations. 

\end{abstract}

\clearpage

\section{Introduction}

Laboratory measurements for the spectra of Br-like ions with 38 $\le$ Z $\le$ 42 have been made by several workers, due to their relevance to studies of fusion plasmas. For example, \cite{tom}--\cite{pw}  observed many lines in the spectra of Sr IV,  and \cite{pr}--\cite{ac} of Y V and Zr VI. Similarly, several lines of Nb VII and Mo VIII have been detected by \cite{msz}--\cite{csk}. These elements, particularly Mo, are used in tokamak reactor walls and  radiation from low charge states of sputtered or evaporated high-Z metal ions, such as Mo VIII, provide a useful study of the spectra \cite{pc}--\cite{vbk}. Measurements of energy levels have been compiled by the NIST (National Institute of Standards and Technology) team \cite{nist}, and are  available at their  website {\tt http://physics.nist.gov/PhysRefData/ASD/levels\_form.html}. However, the corresponding theoretical work on these ions is very limited. Bi{\' e}mont {\em et al} \cite{bch} reported radiative rates (A- values) for the magnetic dipole (M1) and electric quadrupole (E2) transitions among the ground state levels (4s$^2$4p$^5$ $^2$P$^o_{3/2}$ -- $^2$P$^o_{1/2}$) of several Br-like ions, including those considered here. However, such  limited data for a single transition are not particularly useful for modelling applications, with a larger set of results  required. Therefore, in a recent paper Singh {\em at al} \cite{sam} calculated energies for the lowest 31 levels, belonging to  the 4s$^2$4p$^5$, 4s$^2$4p$^4$4d and 4s4p$^6$  configurations, of Br-like ions with 38 $\le$ Z $\le$ 42. Furthermore, they  listed A- values for E1 transitions, but only from the ground state  4s$^2$4p$^5$ $^2$P$^o_{3/2,1/2}$ to higher lying levels,  insufficient for detailed plasma modelling. More importantly, for the calculations they included limited CI (configuration interaction) whereas it is very important for Br-like ions as already demonstrated for another ion, namely W XL \cite{w40a,w40b}.
Therefore, there is scope for improvement as well as an extension to their reported data.

For the calculations we have adopted the {\sc grasp} (general-purpose relativistic atomic structure package) code to generate the wavefunctions. This  was originally developed by Grant  et al.  \cite{grasp0},  has been updated by one of the authors (Dr. P. H. Norrington),  is referred to as  GRASP0 and is freely available at the website {\tt http://web.am.qub.ac.uk/DARC/}. It is a fully relativistic code,  based on the $jj$ coupling scheme. Additional relativistic corrections arising from the Breit interaction and QED (quantum electrodynamics) effects are also included. Furthermore, this version yields comparable results with other revised ones, such as GRASP2K \cite{grasp2k,grasp2kk}.


\section{Energy levels}

Singh {\em at al} \cite{sam} performed two sets of calculations with differing amount of CI. In the first  (GRASP1) they included only three basic configurations, namely 4s$^2$4p$^5$, 4s$^2$4p$^4$4d and 4s4p$^6$, while in the second  (GRASP2a) they added a further five configurations, namely 4s$^2$4p$^4$4f, 4s4p$^5$4d, 4s4p$^5$4f, 4s$^2$4p$^3$4f$^2$ and 4s$^2$4p$^3$4d$^2$. These 8 configurations generate 470 levels in total,  listed in table 1. However, there are many more configurations, such as 4s4p$^5$5$\ell$ and 4s$^2$4p$^4$5$\ell$, whose energy levels closely interact and intermix with the 8 considered by Singh {\em at al}. Based on several tests with increasing amount of CI, we have identified 39 configurations whose energy levels are in the interacting range of energy, and hence are influential in improving the accuracy of the calculations. These configurations and their generated energy ranges are listed in table 1 for all five ions of interest, namely Sr IV, Y V, Zr VI, Nb VII and Mo VIII. We also note that other configurations (such as 3p$^5$3d$^{10}$4s$^2$4p$^6$, 3p$^5$3d$^{10}$4s$^2$4p$^5$4d and 3p$^5$3d$^{10}$4s$^2$4p$^5$4f) have also been included in tests, but their impact on the lower energy levels is insignificant, mainly because they generate levels at much higher energy ranges. For example, their energy range is 19--22 Ryd for Sr IV and 29--36 Ryd for Mo VIII. Nevertheless, we discuss their impact later in the section.

In tables 2--6 we list our energy levels for ions with 38 $\le$ Z $\le$ 42. Included in these tables are our results using two sets of configurations, i.e. 8 (GRASP2b) which are the same as those considered by Singh {\em at al} \cite{sam} and all 39 (GRASP3) listed in table 1. Furthermore, only the lowest 31 levels of the 4s$^2$4p$^5$, 4s$^2$4p$^4$4d and 4s4p$^6$  configurations are included in these tables, because Singh {\em at al} reported results only for these. The corresponding compiled energies of NIST are also provided in the tables to facilitate comparison. Furthermore,  the level orderings  in these tables are the same as those by Singh {\em et al},  even though our calculations and the NIST compilations differ in a few instances -- see for example, levels 29 and 30 of Zr VI in table 4. We also note that the identification of these 31 levels for ions with 38 $\le$ Z $\le$ 42 is not as difficult as was the case for another Br-like ion, i.e. W XL \cite{w40a,w40b}. For this reason we do not provide mixing coefficients for these ions.

The discrepancies between the GRASP2a and GRASP2b energy levels are up to 0.15 Ryd for all ions. Such large discrepancies are {\em not} expected, particularly because  (i) Singh {\em at al} \cite{sam} have adopted the same version of the GRASP code, (ii)  have  used the option of {\em extended average level} (EAL) as by us, and (iii) we are able to reproduce their results corresponding to the GRASP1 calculations for {\em all} ions. In general, their calculated energies are {\em lower} and hence closer to the NIST compilations, although differences are up to $\sim$ 0.2 Ryd (i.e. $\sim$ 7\%) for some levels. However,  we are unable to reproduce their results, and hence have little  confidence in their data. Nevertheless, the limited CI included in these GRASP2 calculations is not sufficient to provide accurate results as already stated in section 1. Therefore, our final results from a much larger calculation (GRASP3) are also listed in tables 2--6. The 39 configurations included in these calculations generate 3990 levels in total.

As a consequence of the increased CI in the GRASP3 calculations, energies for most of the levels have decreased, by up to  $\sim$ 0.2 Ryd for all ions. However,  in comparison to the measurements the calculated energies remain higher, by up to  $\sim$ 0.2 Ryd (i.e. 5\%), depending on the ion. Although we have included a large CI in these calculations, a few other configurations with $n  >$ 5 have been omitted, such as 4p$^6$6$\ell$. Therefore,  to  assess the accuracy of our results, we have also employed the  {\em Flexible Atomic Code} ({\sc fac}) of Gu \cite{fac},  available from the website {\tt http://sprg.ssl.berkeley.edu/$\sim$mfgu/fac/}. This is also a fully relativistic code and provides a variety of atomic parameters. In particular,  results obtained from FAC for energy levels and radiative rates are  comparable to {\sc grasp}, as already shown for several other ions, see for example:  Aggarwal  et al. for Kr \cite{kr} and Xe \cite{xe} ions. Additionally,  larger calculations can  be performed with this code  within a reasonable time frame of a few days. Thus results from {\sc fac} will be helpful in assessing the accuracy of our energy levels and radiative rates.

As with {\sc grasp}, we have also performed a series of calculations  with the {\sc fac} code with increasing amount of CI,  but focus only on three. These are (i) FAC1, which includes the same 470 levels as in GRASP2, (ii) FAC2,   which includes the same 3990 levels as in GRASP3,  and finally (iii) FAC3, which  includes a total of 12,137 levels, the additional ones arising from the 4p$^6$6$\ell$,  4s4p$^5$6$\ell$, 4s$^2$4p$^4$6$\ell$,   4p$^6$7$\ell$,  4s4p$^5$7$\ell$, 4s$^2$4p$^4$7$\ell$, 4s$^2$4p$^3$5$\ell^2$, 4s4p$^4$5$\ell^2$, 3p$^5$3d$^{10}$4s$^2$4p$^6$, 3p$^5$3d$^{10}$4s$^2$4p$^5$4d and 3p$^5$3d$^{10}$4s$^2$4p$^5$4f configurations. Results obtained from all  three calculations are also listed in tables 2--6.

With the same levels of CI (i.e. GRASP2b and FAC1 and GRASP3 and FAC2) both sets of energies from two independent codes agree within 0.1 Ryd for all levels and all ions listed in tables 2--6. This is highly satisfactory and confirms, once again, that the energies reported by Singh {\em at al} \cite{sam} are difficult to understand. Energies obtained from calculations with larger CI (i.e. FAC3) agree with those from FAC2 to within 0.01 Ryd for all levels and ions. This clearly indicates that the CI included in our GRASP3 and FAC2 calculations is more than sufficient, and hence there is no need to include additional CI as far as the levels of the 4s$^2$4p$^5$, 4s$^2$4p$^4$4d and 4s4p$^6$  configurations are concerned. In conclusion, all theoretical energies are higher than the NIST compilations and results obtained from our GRASP3 calculations are the closest. For most levels, the discrepancies between theory and measurements are small, but the differences are larger (up to $\sim$ 0.2 Ryd or 5\%) for some of the higher levels and hence this is also our assessment of the accuracy.

\section{Radiative rates}

Besides energy levels, radiative rates (A- values) are required for the modelling of plasmas. Generally, electric dipole (E1) transitions dominate in magnitude, but sometimes other types of transitions are also important and hence are preferred for  a  complete plasma model. Their availability is more important for the further calculations of lifetimes -- see section 4. Therefore, we have also calculated A- values for the electric quadrupole (E2), magnetic dipole (M1) and  magnetic quadrupole (M2) transitions.  For all types of transitions, the absorption oscillator strength ($f_{ij}$) and radiative rate A$_{ji}$ (in s$^{-1}$)  are related by the following expression:

\begin{equation}
f_{ij} = \frac{mc}{8{\pi}^2{e^2}}{\lambda^2_{ji}} \frac{{\omega}_j}{{\omega}_i} A_{ji}
 = 1.49 \times 10^{-16} \lambda^2_{ji}    \frac{{ \omega}_j}{{\omega}_i} A_{ji}                                 
\end{equation}
where $m$ and $e$ are the electron mass and charge, respectively, $c$  the velocity of light,  $\lambda_{ji}$  the transition energy/wavelength in $\rm \AA$, and $\omega_i$ and $\omega_j$  the statistical weights of the lower $i$ and upper $j$ levels, respectively. Similarly, if required, corresponding S- values can be obtained through Eqs. (2--5) of  \cite{tixix}.

 The A-,  f-  and S- values have been calculated in both Babushkin and Coulomb gauges, i.e.  the length and velocity forms in the widely used non-relativistic nomenclature. However,  the velocity form is generally  considered to be comparatively less accurate and therefore we will present results in the length form alone, as did Singh {\em at al} \cite{sam}.  Nevertheless,  we will discuss  later the velocity/length  form ratio, as this provides some assessment of  the accuracy of the results. 
 
 In tables 7--11 we compare our A- values from two calculations with the GRASP code (GRASP2b and GRASP3) with those of GRASP2a \cite{sam} for ions with 38 $\le$ Z $\le$ 42. Also included in these tables are the f- values from our GRASP3 calculations, because they give an indication of the strength of a transition. Only transitions from the ground state levels 4s$^2$4p$^5$ $^2$P$^o_{3/2,1/2}$ are included as these are the ones reported by Singh {\em at al} \cite{sam}. As for the energy levels, we cannot reproduce their A- values with the configurations they  included. Differences of a factor of two between the two sets of A-  values are common for many transitions and all ions -- see for example, the 1--3/19 and  2--3/6/27 transitions of Sr IV in table 7. For some transitions, their A- values are lower whereas  others  are higher, and hence  there is no systematic trend in the differences. Similarly, for a few transitions, discrepancies between the two sets of A- values are up to two orders of magnitude -- see for example, 2--6 of Y V,  Zr VI and Nb VII, in tables 8--10. However,  for the 1--26 transition of Zr VI  discrepancy is up to five orders of magnitude. Some of these transitions are {\em weak} (f $\sim$ 10$^{-4}$ or less) and therefore some differences in the A- values are not unexpected, but not of  orders of magnitude, particularly when the same atomic structure code and the same level of CI are used. Therefore, we do not discuss the A- values of Singh {\em at al} further and rather focus on the accuracy assessment of our results.

As noted in section 1 and discussed in section 2, the CI included in the GRASP2 calculations is insufficient for accurate determination of atomic parameters. Therefore, results obtained from our larger GRASP3 calculations should be comparatively more accurate, as for the energy levels. Differences between the GRASP2b and GRASP3 A- values are up to  a factor of two for several transitions and in all ions -- see for example, 1--5/7/10/11 of Sr IV in table 7. However, all such transitions have small f- values, and are  more susceptible to varying amount of CI, due to cancellation or additive effects. Therefore,  such discrepancies are very common  \cite{oh}. For the same reason, differences between the two sets of A- values are larger (up to an order of magnitude) for some weaker transitions, such as 1--26 (f = 5.5$\times$10$^{-2}$), 2--6 (f = 2.6$\times$10$^{-5}$) and 2--14 (f = 5.5$\times$10$^{-5}$). Finally, discrepancies, if any, with the Bi{\' e}mont {\em et al} \cite{bch} A- values for the M1 and E2 (4s$^2$4p$^5$) $^2$P$^o_{3/2}$ -- $^2$P$^o_{1/2}$ transitions  are less than $\sim$ 10\% for all ions.

Also included in tables 7--11 are the ratio (R) of the velocity and length forms of A- values. For strong transitions (f $\ge$ 0.01) R is within $\sim$20\% of unity, which is highly satisfactory. However, the 1--3 and 2--3 (i.e. 4s$^2$4p$^5$ $^2$P$^o_{3/2,1/2}$ -- 4s4p$^6$ $^2$S$_{1/2}$) transitions are exceptions, because R is up to 0.5 depending on the ion. For these transitions CI is not very influential because the A- values from GRASP2b and GRASP3 are similar. Such discrepancies are often observed for even comparatively strong transitions. Moreover, different calculations with varying amount of CI may yield R closer to unity but strikingly  different f- values in magnitude -- see Aggarwal {\em et al} \cite{fe15} for examples. Among the weaker transitions, discrepancies are up to an order of magnitude for a few, and the only exception is the 2--6 transition (f $\sim$ 10$^{-11}$) of Zr VI for which R = 1200. Therefore, based on the ratio R and other comparisons discussed above,  the accuracy of our listed results is assessed to be better than 20\% for a majority of the transitions.

Since no other results of comparable complexity and accuracy are available for the A- values, we have undertaken calculations with FAC, because in general it generates results of comparable accuracy \cite{fe15}.  For brevity, we include in  Table 7 the A-  values for transitions of Sr IV obtained from the FAC2  calculations, which include the same CI (3990 levels) as in GRASP3. Corresponding results from a larger calculation with 12,137 levels (FAC3) are also included in the table. For most (comparatively strong transitions with f $\ge$ 0.01) the GRASP3 and FAC2 A- values agree within $\sim$20\%, although differences of up to 50\% are also noted for a few, such as 1--25/26 and 2--27. Differences are larger, up to a factor of 5, for a few weaker transitions -- see for example  1--12 (f $\sim$ 10$^{-5}$). Such differences between two independent calculations are not unexpected and have been noted in the past for several ions. Additionally, for the 1--14/15 transitions the A- values in FAC2 (and FAC3) appear to be interchanged, but not for 2--14/15.  The effect of larger CI can be better understood with the comparison between the FAC2 and FAC3 A- values. For some weaker transitions, differences are again up to a factor of 5, and are up to 50\% for a few stronger ones, such as 1--25/26/30  and 2--27. Surprisingly, for three of  these transitions (except 1--30) the A- values from FAC3 are closer to GRASP3 than to FAC2. Similar differences between calculations with the {\sc grasp}  and {\sc fac}  codes are noted for transitions of other ions. Therefore, we have confidence in our GRASP3 calculations, and as already stated our accuracy assessment for a majority of (strong) transitions remains to be about 20\%, although scope remains for improvement.  

\section{Lifetimes}

The lifetime $\tau$ of a level $j$ is defined as follows:

\begin{equation}
{\tau}_j = \frac{1}{{\sum_{i}^{}} A_{ji}}.
\end{equation}

In tables 12--16 we have  listed lifetimes for all 31 levels of the 38 $\le$ Z $\le$ 42 ions. Included in these tables are our results from the GRASP2b and GRASP3 calculations along with those of Singh {\em at al} \cite{sam}, designated as GRASP2a. For our calculations, the  contribution of all types of transitions (E1, E2, M1 and M2) are included. Additionally, we have also listed the A- values for the {\em dominant} transitions, i.e. all those which have a contribution of $\ge$ 20\% to $\sum$A$_{ji}$.

Unfortunately, the only other results for $\tau$ available for comparison are those of Singh {\em at al} \cite{sam}, who have also included contributions from all four types of transitions, and have adopted the same version of the GRASP code as already stated. Their results are included in tables 12--16, but discrepancies with our calculations are up to four orders of magnitude for all ions  -- see for example, levels 24 of Sr IV, 20 of Y V, 26 of Zr VI, and 5 of Nb VII and Mo VIII. However, as the A- values of Singh {\em at al}  are not considered to be reliable, as discussed in section 3, neither are the results for $\tau$.

Since our GRASP3 calculations include larger CI, and the energy levels and A- values from these are comparatively more accurate, so should be the corresponding results for $\tau$. Differences between GRASP2b and GRASP3 values of $\tau$ are generally within a factor of two, although discrepancies for a few levels, such as 10 of Sr IV and 26 of Y V, are up to two orders of magnitude. However, the dominant  transitions for all such levels are invariably weak and hence less accurate, as already discussed in section 3. Further calculations and preferably future measurements of $\tau$ will be more helpful in assessing the accuracy of our results.

\section{Conclusions}

In this work, energy levels, radiative rates, oscillator strengths and lifetimes have been listed for transitions among the lowest 31 levels of Br-like ions with 38 $\le$ Z $\le$ 42. These levels belong to the  4s$^2$4p$^5$, 4s$^2$4p$^4$4d and 4s4p$^6$  configurations, and for calculations the GRASP code has been adopted.  Extensive comparisons have been made for all atomic parameters with the differing amount of CI as well as with similar calculations with the FAC code. Unfortunately the only prior results available for comparison are those of Singh {\em at al} \cite{sam}, which are not re-produceable.  Although differences  with their energy levels are only up to 0.15 Ryd, the discrepancies for radiative rates and lifetimes are up to five orders of magnitude for several transitions (levels) and for all ions. 

Differences between our energy levels calculated with a large CI and the measurements (compiled by NIST) are up to  0.2 Ryd  ($\sim$ 5\%) for a few, although the agreement is much better for most levels. Similarly, the accuracy for other parameters, namely A-, f- and $\tau$ values, is assessed to be better than 20\% for a majority of (strong) transitions/levels. Finally,  calculations for energies  and A- values have been performed for up to 12,137  levels, although results have been discussed for only 31.  Results for a larger number of levels and their corresponding transitions  will be reported in a later paper.

\section*{Acknowledgment}
 KMA  is thankful to  AWE Aldermaston for financial support.     


\newpage
\clearpage

\begin{flushleft}
Table 1. Configurations of Br-like ions, their generated levels and energy ranges (in Ryd).
\end{flushleft}
{\small
\begin{tabular}{rllrrrrrrrrrr} \hline
 & & & & & & & &  \\
Index  & Configuration         & No. of Levels               & Sr IV & Y V & Zr VI & Nb VII & Mo VIII   \\
& & & & & & & &   \\ \hline  
& & & & & & & &   \\
  1  & 4s$^2$4p$^5$     	          & 2$^{\rm o}$     &	 0.0  -- 0.1	& 0.0  --  0.1  & 0.0  -- 0.1	  & 0.0   --  0.2     & 0.0    --  0.2     \\
  2  & 4s$^2$4p$^4$4d		  	  & 28  	    &	 1.8  -- 2.6	& 2.1  --  3.0  & 2.4  -- 3.5	  & 2.7   --  3.9     & 3.0    --  4.3     \\
  3  & 4s$^2$4p$^4$4f		  	  & 30$^{\rm o}$    &	 3.1  -- 3.5	& 3.8  --  4.3  & 4.6  -- 5.2	  & 5.2   --  5.9     & 5.8    --  6.9     \\
  4  & 4s4p$^6$	      		  	  & 1		    &	 1.5  -- 1.5	& 1.7  --  1.7  & 1.9  -- 1.9	  & 2.1   --  2.1     & 2.3    --  2.3     \\
  5  & 4p$^6$4d	      		  	  & 2		    &	 5.5  -- 5.5	& 6.4  --  6.4  & 6.9  -- 7.2	  & 7.5   --  8.2     & 8.2    --  9.1     \\
  6  & 4p$^6$4f	      		  	  & 2$^{\rm o}$     &	 6.9  -- 7.0	& 8.1  --  8.4  & 9.3  -- 9.6	  & 10.3  --   10.8   & 11.8   --   12.1   \\
  7  & 4s4p$^5$4d		  	  & 23$^{\rm o}$    &	 3.1  -- 3.6	& 3.6  --  4.3  & 4.0  -- 5.1	  & 4.5   --  5.7     & 5.1    --  5.9     \\
  8  & 4s4p$^5$4f		  	  & 24  	    &	 4.6  -- 5.7	& 5.5  --  6.9  & 6.5  -- 8.0	  & 7.4   --  8.1     & 8.3    --  9.0     \\
  9  & 4s$^2$4p$^3$4d$^2$	  	  & 141$^{\rm o}$   &	 3.6  -- 5.0	& 4.1  --  5.7  & 4.7  -- 6.6	  & 5.2   --  7.4     & 5.8    --  8.1     \\
 10  & 4s$^2$4p$^3$4f$^2$	  	  & 221$^{\rm o}$   &	 6.5  -- 7.2	& 8.0  --  8.6  & 9.4  -- 10.3    & 10.7  --   11.8   & 12.0   --   13.6   \\
 11  & 4s$^2$4p$^3$4d4f		  	  & 363 	    &	 4.9  -- 6.1	& 6.0  --  7.3  & 7.0  -- 8.6	  & 8.0   --  9.9     & 8.9    --  11.4    \\
 12  & 4s$^2$4p$^2$4d$^3$	  	  & 261 	    &	 5.7  -- 7.3	& 6.5  --  8.6  & 7.4  -- 9.8	  & 8.1   --  11.0    & 9.0    --  12.2    \\
 13  & 4s$^2$4p4d$^4$		  	  & 180$^{\rm o}$   &	 8.2  -- 9.7	& 9.5  --  11.3 & 10.8 --  12.9   & 11.8  --   14.4   & 13.1   --   16.0   \\
 14  & 4s$^2$4p$^2$4d$^2$4f	  	  & 1140$^{\rm o}$  &	 7.0  -- 8.5	& 8.3  --  10.2 & 9.5  -- 11.6    & 10.7  --   13.4   & 11.9   --   15.2   \\
 15  & 4s4p$^3$4d$^3$    	  	  & 678$^{\rm o}$   &	 6.5  -- 9.2	& 7.4  --  10.7 & 8.4  -- 12.2    & 9.3   --  13.7    & 10.3   --   15.1   \\
 16  & 4p$^5$4d$^2$		  	  & 45$^{\rm o}$    &	 6.8  -- 7.5	& 7.7  --  8.4  & 8.7  -- 9.6	  & 9.6   --  10.8    & 10.6   --   11.8   \\
 17  & 3d$^9$4s$^2$4p$^5$4d	  	  & 96$^{\rm o}$    &	 10.3 -- 11.4	& 12.1 --  13.4 & 13.9 --  15.5   & 15.9  --   17.7   & 17.9   --   19.9   \\
 18  & 3d$^9$4s$^2$4p$^5$4f	  	  & 113 	    &	 11.8 -- 12.1	& 14.1 --  14.5 & 16.4 --  16.9   & 18.8  --   19.5   & 21.2   --   22.1   \\
 19  & 3d$^9$4s$^2$4p$^6$	  	  & 2		    &	 8.4  -- 8.6	& 9.9  --  10.1 & 11.5 --  11.7   & 13.2  --   13.4   & 15.0   --   15.2   \\
 20  & 4s4p$^5$5s		  	  & 7$^{\rm o}$     &	 3.8  -- 4.4	& 4.6  --  5.3  & 5.3  -- 6.2	  & 6.3   --  7.2     & 7.3    --  8.2     \\
 21  & 4s4p$^5$5p		  	  & 18  	    &	 4.1  -- 4.8	& 5.0  --  5.8  & 5.9  -- 6.8	  & 6.9   --  7.8     & 7.9    --  9.0     \\
 22  & 4s4p$^5$5d		  	  & 23$^{\rm o}$    &	 4.6  -- 5.3	& 5.7  --  6.4  & 6.8  -- 7.6	  & 7.9   --  8.8     & 9.1    --  10.1    \\
 23  & 4s4p$^5$5f		  	  & 24  	    &	 5.1  -- 5.7	& 6.3  --  7.0  & 7.5  -- 8.3	  & 8.8   --  9.7     & 10.2   --   11.1   \\
 24  & 4s4p$^5$5g		  	  & 24$^{\rm o}$    &	 5.1  -- 5.7	& 6.4  --  7.0  & 7.7  -- 8.4	  & 9.1   --  9.9     & 10.5   --   11.4   \\
 25  & 4p$^6$5s 		  	  & 1		    &	 6.0  -- 6.0	& 7.1  --  7.1  & 8.2  -- 8.2	  & 9.3   --  9.3     & 10.5   --   10.5   \\
 26  & 4p$^6$5p 		  	  & 2$^{\rm o}$     &	 6.3  -- 6.3	& 7.4  --  7.6  & 8.7  -- 8.7	  & 9.8   --  10.0    & 11.1   --   11.2   \\
 27  & 4p$^6$5d 		  	  & 2		    &	 6.9  -- 6.9	& 8.1  --  8.1  & 9.5  -- 9.5	  & 10.9  --   10.9   & 12.3   --   12.3   \\
 28  & 4p$^6$5f 		  	  & 2$^{\rm o}$     &	 7.2  -- 7.3	& 8.7  --  8.7  & 10.2 --  10.3   & 11.7  --   11.8   & 13.4   --   13.5   \\
 29  & 4p$^6$5g 		  	  & 2		    &	 7.3  -- 7.3	& 8.8  --  8.8  & 10.4 --  10.4   & 12.0  --   12.0   & 13.7   --   13.7   \\
 30  & 4s$^2$4p$^4$5s		  	  & 8		    &	 2.2  -- 2.6	& 2.8  --  3.3  & 3.5  -- 4.0	  & 4.2   --  4.8     & 4.9    --  5.6     \\
 31  & 4s$^2$4p$^4$5p		  	  & 21$^{\rm o}$    &	 2.5  -- 2.9	& 3.2  --  3.7  & 4.0  -- 4.6	  & 4.8   --  5.4     & 5.6    --  6.4     \\
 32  & 4s$^2$4p$^4$5d		  	  & 28  	    &	 3.0  -- 3.5	& 3.9  --  4.4  & 4.8  -- 5.4	  & 5.8   --  6.4     & 6.8    --  7.5     \\
 33  & 4s$^2$4p$^4$5f		  	  & 30$^{\rm o}$    &	 3.5  -- 3.9	& 4.5  --  4.9  & 5.4  -- 6.2	  & 6.7   --  7.4     & 7.8    --  8.6     \\
 34  & 4s$^2$4p$^4$5g		  	  & 30  	    &	 3.5  -- 3.9	& 4.6  --  5.0  & 5.7  -- 6.2	  & 6.9   --  7.5     & 8.1    --  8.8     \\
 35  & 3d$^9$4s$^2$4p$^5$5s	  	  & 23$^{\rm o}$    &	 10.8 -- 11.2	& 12.9 --  13.3 & 15.2 --  15.7   & 17.6  --   18.1   & 20.1   --   20.7   \\
 36  & 3d$^9$4s$^2$4p$^5$5p	  	  & 65  	    &	 11.1 -- 11.7	& 13.3 --  14.0 & 15.7 --  16.4   & 18.2  --   19.0   & 20.8   --   21.7   \\
 37  & 3d$^9$4s$^2$4p$^5$5d	  	  & 96$^{\rm o}$    &	 11.7 -- 12.1	& 14.1 --  14.6 & 16.6 --  17.2   & 19.2  --   19.9   & 22.0   --   22.8   \\
 38  & 3d$^9$4s$^2$4p$^5$5f	  	  & 113 	    &	 12.2 -- 12.5	& 14.7 --  15.1 & 17.4 --  17.8   & 20.2  --   20.7   & 23.1   --   23.7   \\
 39  & 3d$^9$4s$^2$4p$^5$5g	  	  & 119$^{\rm o}$   &	 12.2 -- 12.5	& 14.8 --  15.2 & 17.5 --  18.0   & 20.4  --   20.9   & 23.4   --   24.0   \\
& &  &  & & & & & & \\ \hline	   			   		   
\end{tabular}						   																							
\begin{flushleft}													       
{\small
																       
}															       
\end{flushleft} 
\newpage
\clearpage

\begin{flushleft}
Table 2. Energies (Ryd) for the lowest 31 levels of Sr IV. 
\end{flushleft}
{\small
\begin{tabular}{rllrrrrrrrrrr} \hline
 & & & & & & & &  \\
Index  & Configuration         & Level               & NIST     & GRASP2a & GRASP2b & GRASP3  &  FAC1	& FAC2     &  FAC3   \\
& & & & & & & &   \\ \hline  
& & & & & & & &   \\
  1  & 4s$^2$4p$^5$	       &  $^2$P$^o_{3/2}$    & 0.00000  & 0.00000 & 0.00000 & 0.00000 & 0.00000 & 0.00000  & 0.00000 \\
  2  & 4s$^2$4p$^5$	       &  $^2$P$^o_{1/2}$    & 0.08865  & 0.08580 & 0.08396 & 0.08510 & 0.09128 & 0.08801  & 0.08812 \\
  3  & 4s4p$^6$ 	       &  $^2$S$  _{1/2}$    & 1.37149  & 1.34619 & 1.42587 & 1.44410 & 1.46260 & 1.47424  & 1.46391 \\
  4  & 4s$^2$4p$^4$($^3$P)4d   &  $^4$D$  _{7/2}$    & 1.70553  & 1.65949 & 1.72016 & 1.75580 & 1.77666 & 1.79194  & 1.78496 \\
  5  & 4s$^2$4p$^4$($^3$P)4d   &  $^4$D$  _{5/2}$    & 1.70590  & 1.66119 & 1.72216 & 1.75740 & 1.77668 & 1.79181  & 1.78502 \\
  6  & 4s$^2$4p$^4$($^3$P)4d   &  $^4$D$  _{3/2}$    & 1.71340  & 1.66959 & 1.73051 & 1.76560 & 1.78486 & 1.79993  & 1.79324 \\
  7  & 4s$^2$4p$^4$($^3$P)4d   &  $^4$D$  _{1/2}$    & 1.72339  & 1.67979 & 1.74053 & 1.77570 & 1.79724 & 1.81238  & 1.80563 \\
  8  & 4s$^2$4p$^4$($^3$P)4d   &  $^4$F$  _{9/2}$    & 1.79574  & 1.76139 & 1.82292 & 1.85900 & 1.86623 & 1.88674  & 1.88277 \\
  9  & 4s$^2$4p$^4$($^3$P)4d   &  $^4$F$  _{7/2}$    & 1.82563  & 1.79409 & 1.85556 & 1.89120 & 1.89699 & 1.91794  & 1.91435 \\
 10  & 4s$^2$4p$^4$($^1$D)4d   &  $^2$P$  _{1/2}$    & 1.82736  & 1.81929 & 1.88463 & 1.92440 & 1.92252 & 1.94584  & 1.94315 \\
 11  & 4s$^2$4p$^4$($^3$P)4d   &  $^4$F$  _{5/2}$    & 1.85301  & 1.81889 & 1.87983 & 1.91360 & 1.92612 & 1.94461  & 1.94092 \\
 12  & 4s$^2$4p$^4$($^3$P)4d   &  $^4$F$  _{3/2}$    & 1.86062  & 1.82929 & 1.89037 & 1.92180 & 1.93655 & 1.95160  & 1.94816 \\
 13  & 4s$^2$4p$^4$($^3$P)4d   &  $^4$P$  _{1/2}$    & 1.86576  & 1.84949 & 1.91510 & 1.95600 & 1.94719 & 1.97585  & 1.97497 \\
 14  & 4s$^2$4p$^4$($^3$P)4d   &  $^4$P$  _{3/2}$    & 1.86630  & 1.85239 & 1.91773 & 1.95620 & 1.95263 & 1.97723  & 1.97562 \\
 15  & 4s$^2$4p$^4$($^1$D)4d   &  $^2$D$  _{3/2}$    & 1.88199  & 1.87389 & 1.93923 & 1.97160 & 1.97384 & 1.99236  & 1.99005 \\
 16  & 4s$^2$4p$^4$($^3$P)4d   &  $^2$F$  _{7/2}$    & 1.89068  & 1.87849 & 1.94197 & 1.98050 & 1.97339 & 1.99994  & 1.99863 \\
 17  & 4s$^2$4p$^4$($^3$P)4d   &  $^4$P$  _{5/2}$    & 1.90398  & 1.88809 & 1.95321 & 1.98860 & 1.98671 & 2.00978  & 2.00848 \\
 18  & 4s$^2$4p$^4$($^1$D)4d   &  $^2$P$  _{3/2}$    & 1.90647  & 1.89589 & 1.96084 & 1.99670 & 1.99774 & 2.01961  & 2.01784 \\
 19  & 4s$^2$4p$^4$($^1$D)4d   &  $^2$D$  _{5/2}$    & 1.93164  & 1.92249 & 1.98705 & 2.02010 & 2.01960 & 2.04090  & 2.03959 \\
 20  & 4s$^2$4p$^4$($^3$P)4d   &  $^2$F$  _{5/2}$    & 1.95873  & 1.95119 & 2.01635 & 2.05440 & 2.04978 & 2.07643  & 2.07603 \\
 21  & 4s$^2$4p$^4$($^1$D)4d   &  $^2$G$  _{9/2}$    & 1.96093  & 1.95249 & 2.01370 & 2.04620 & 2.04955 & 2.06689  & 2.06394 \\
 22  & 4s$^2$4p$^4$($^1$D)4d   &  $^2$G$  _{7/2}$    & 1.96094  & 1.95399 & 2.01703 & 2.05040 & 2.05268 & 2.07196  & 2.06975 \\
 23  & 4s$^2$4p$^4$($^1$D)4d   &  $^2$F$  _{5/2}$    & 2.05829  & 2.07879 & 2.14641 & 2.18590 & 2.16103 & 2.19311  & 2.19528 \\
 24  & 4s$^2$4p$^4$($^1$D)4d   &  $^2$F$  _{7/2}$    & 2.07858  & 2.09719 & 2.16432 & 2.20420 & 2.17931 & 2.21245  & 2.21470 \\
 25  & 4s$^2$4p$^4$($^1$S)4d   &  $^2$D$  _{3/2}$    & 2.20683  & 2.26209 & 2.33287 & 2.29400 & 2.33357 & 2.30394  & 2.30686 \\
 26  & 4s$^2$4p$^4$($^1$S)4d   &  $^2$D$  _{5/2}$    & 2.23930  & 2.28919 & 2.35916 & 2.31430 & 2.36409 & 2.32614  & 2.32847 \\
 27  & 4s$^2$4p$^4$($^3$P)4d   &  $^2$P$  _{3/2}$    & 2.29991  & 2.42829 & 2.49870 & 2.41760 & 2.47141 & 2.45258  & 2.45337 \\
 28  & 4s$^2$4p$^4$($^3$P)4d   &  $^2$P$  _{1/2}$    & 2.30762  & 2.45749 & 2.52825 & 2.44470 & 2.50013 & 2.47961  & 2.48054 \\
 29  & 4s$^2$4p$^4$($^3$P)4d   &  $^2$D$  _{5/2}$    & 2.32210  & 2.48659 & 2.55533 & 2.45350 & 2.52189 & 2.49083  & 2.48627 \\
 30  & 4s$^2$4p$^4$($^1$D)4d   &  $^2$S$  _{1/2}$    & 2.34510  & 2.54509 & 2.62876 & 2.51540 & 2.57417 & 2.54771  & 2.52024 \\
 31  & 4s$^2$4p$^4$($^3$P)4d   &  $^2$D$  _{3/2}$    & 2.40740  & 2.56949 & 2.63736 & 2.52380 & 2.61111 & 2.56987  & 2.56654 \\
& & & & & & & & \\ \hline            								                	 
\end{tabular}   								   					       
			      							   					       
\begin{flushleft}													       
{\small
NIST: http://physics.nist.gov/PhysRefData/ASD/levels\_form.html \\
GRASP2a: Singh {\em et al} \cite{sam} \\ 
GRASP2b: present calculations from the {\sc grasp} code with 470 levels\\
GRASP3: present calculations from the {\sc grasp} code with 3990 levels\\
FAC1: present calculations from the {\sc fac} code with 470 levels  \\  
FAC2: present calculations from the {\sc fac} code with 3990 levels \\ 													
FAC3: present calculations from the {\sc fac} code with 12,137 levels \\																       
}															       
\end{flushleft} 

\newpage
\clearpage

\begin{flushleft}
Table 3. Energies (Ryd) for the lowest 31 levels of Y V. 
\end{flushleft}
{\small
\begin{tabular}{rllrrrrrrrrrr} \hline
 & & & & & & & &  \\
Index  & Configuration         & Level               & NIST     & GRASP2a & GRASP2b & GRASP3  &  FAC1	& FAC2    &  FAC3    \\
& & & & & & & &   \\ \hline  
& & & & & & & &   \\
  1  & 4s$^2$4p$^5$	       &  $^2$P$^o_{3/2}$    & 0.00000  & 0.00000 & 0.00000 & 0.00000 & 0.00000 & 0.00000 &  0.00000  \\
  2  & 4s$^2$4p$^5$	       &  $^2$P$^o_{1/2}$    & 0.11354  & 0.11070 & 0.10869 & 0.11016 & 0.11584 & 0.11221 &  0.11242  \\
  3  & 4s4p$^6$ 	       &  $^2$S$  _{1/2}$    & 1.55777  & 1.53529 & 1.63922 & 1.63952 & 1.66378 & 1.68518 &  1.67802  \\
  4  & 4s$^2$4p$^4$($^3$P)4d   &  $^4$D$  _{5/2}$    & 1.99039  & 1.95439 & 2.04350 & 2.03778 & 2.08415 & 2.09809 &  2.09281  \\
  5  & 4s$^2$4p$^4$($^3$P)4d   &  $^4$D$  _{7/2}$    &          & 1.95489 & 2.04279 & 2.03768 & 2.08521 & 2.09935 &  2.09386  \\
  6  & 4s$^2$4p$^4$($^3$P)4d   &  $^4$D$  _{3/2}$    & 1.99909  & 1.96479 & 2.05347 & 2.04756 & 2.09388 & 2.10777 &  2.10266  \\
  7  & 4s$^2$4p$^4$($^3$P)4d   &  $^4$D$  _{1/2}$    &          & 1.97829 & 2.06682 & 2.06129 & 2.10923 & 2.12333 &  2.11821  \\
  8  & 4s$^2$4p$^4$($^3$P)4d   &  $^4$F$  _{9/2}$    &          & 2.06969 & 2.15829 & 2.14949 & 2.18962 & 2.20684 &  2.20336  \\
  9  & 4s$^2$4p$^4$($^3$P)4d   &  $^4$F$  _{7/2}$    &          & 2.10979 & 2.19842 & 2.18916 & 2.22817 & 2.24589 &  2.24292  \\
 10  & 4s$^2$4p$^4$($^1$D)4d   &  $^2$P$  _{1/2}$    & 2.12979  & 2.13549 & 2.22760 & 2.22078 & 2.25419 & 2.27477 &  2.27360  \\
 11  & 4s$^2$4p$^4$($^3$P)4d   &  $^4$F$  _{5/2}$    & 2.16587  & 2.14179 & 2.22998 & 2.21897 & 2.26452 & 2.27915 &  2.27610  \\
 12  & 4s$^2$4p$^4$($^3$P)4d   &  $^4$F$  _{3/2}$    & 2.17077  & 2.15269 & 2.24117 & 2.22643 & 2.27452 & 2.28432 &  2.28175  \\
 13  & 4s$^2$4p$^4$($^3$P)4d   &  $^4$P$  _{1/2}$    & 2.17700  & 2.17569 & 2.26792 & 2.26118 & 2.28892 & 2.31553 &  2.31599  \\
 14  & 4s$^2$4p$^4$($^3$P)4d   &  $^4$P$  _{3/2}$    & 2.17913  & 2.17809 & 2.27008 & 2.26109 & 2.29418 & 2.31587 &  2.31561  \\
 15  & 4s$^2$4p$^4$($^1$D)4d   &  $^2$D$  _{3/2}$    & 2.19568  & 2.19999 & 2.29201 & 2.27849 & 2.31558 & 2.33235 &  2.33154  \\
 16  & 4s$^2$4p$^4$($^3$P)4d   &  $^2$F$  _{7/2}$    &          & 2.20779 & 2.29778 & 2.28741 & 2.32020 & 2.34148 &  2.34040  \\
 17  & 4s$^2$4p$^4$($^3$P)4d   &  $^4$P$  _{5/2}$    & 2.22633  & 2.22529 & 2.31702 & 2.30421 & 2.33938 & 2.35905 &  2.35960  \\
 18  & 4s$^2$4p$^4$($^1$D)4d   &  $^2$P$  _{3/2}$    & 2.22907  & 2.23199 & 2.32365 & 2.31417 & 2.34982 & 2.37023 &  2.37017  \\
 19  & 4s$^2$4p$^4$($^1$D)4d   &  $^2$D$  _{5/2}$    & 2.25868  & 2.26259 & 2.35390 & 2.34111 & 2.37687 & 2.39602 &  2.39615  \\
 20  & 4s$^2$4p$^4$($^1$D)4d   &  $^2$G$  _{9/2}$    &          & 2.28759 & 2.37680 & 2.36325 & 2.40286 & 2.41598 &  2.41363  \\
 21  & 4s$^2$4p$^4$($^1$D)4d   &  $^2$G$  _{7/2}$    &          & 2.29049 & 2.38026 & 2.36683 & 2.40545 & 2.42076 &  2.41929  \\
 22  & 4s$^2$4p$^4$($^3$P)4d   &  $^2$F$  _{5/2}$    & 2.29096  & 2.29989 & 2.39033 & 2.37964 & 2.41329 & 2.43587 &  2.43625  \\
 23  & 4s$^2$4p$^4$($^1$D)4d   &  $^2$F$  _{5/2}$    & 2.39830  & 2.43559 & 2.52895 & 2.51624 & 2.53536 & 2.56282 &  2.56557  \\
 24  & 4s$^2$4p$^4$($^1$D)4d   &  $^2$F$  _{7/2}$    &          & 2.46109 & 2.55404 & 2.54187 & 2.56126 & 2.58950 &  2.59244  \\
 25  & 4s$^2$4p$^4$($^1$S)4d   &  $^2$D$  _{3/2}$    & 2.56288  & 2.63429 & 2.73026 & 2.63982 & 2.72567 & 2.68386 &  2.68757  \\
 26  & 4s$^2$4p$^4$($^1$S)4d   &  $^2$D$  _{5/2}$    & 2.60623  & 2.67089 & 2.76600 & 2.67046 & 2.76567 & 2.71562 &  2.71925  \\
 27  & 4s$^2$4p$^4$($^3$P)4d   &  $^2$P$  _{3/2}$    & 2.70710  & 2.85948 & 2.95379 & 2.80873 & 2.92552 & 2.88542 &  2.82960  \\
 28  & 4s$^2$4p$^4$($^3$P)4d   &  $^2$P$  _{1/2}$    & 2.75807  & 2.88958 & 2.98556 & 2.84194 & 2.95321 & 2.91641 &  2.91254  \\
 29  & 4s$^2$4p$^4$($^3$P)4d   &  $^2$D$  _{5/2}$    & 2.73581  & 2.92628 & 3.01877 & 2.86031 & 2.98561 & 2.94256 &  2.93524  \\
 30  & 4s$^2$4p$^4$($^1$D)4d   &  $^2$S$  _{1/2}$    & 2.68791  & 2.94618 & 3.04846 & 2.93707 & 3.00536 & 2.99160 &  2.99604  \\
 31  & 4s$^2$4p$^4$($^3$P)4d   &  $^2$D$  _{3/2}$    & 2.84354  & 3.02998 & 3.12163 & 2.95577 & 3.09521 & 3.03878 &  3.03076  \\
& & & & & & & & \\ \hline            								                	 
\end{tabular}   								   					       
			      							   					       
\begin{flushleft}													       
{\small
NIST: http://physics.nist.gov/PhysRefData/ASD/levels\_form.html \\
GRASP2a: Singh {\em et al} \cite{sam} \\ 
GRASP2b: present calculations from the {\sc grasp} code with 470 levels\\
GRASP3: present calculations from the {\sc grasp} code with 3990 levels\\
FAC1: present calculations from the {\sc fac} code with 470 levels  \\  
FAC2: present calculations from the {\sc fac} code with 3990 levels \\ 													
FAC3: present calculations from the {\sc fac} code with 12,137 levels \\																       
}															       
\end{flushleft} 

\newpage
\clearpage

\begin{flushleft}
Table 4. Energies (Ryd) for the lowest 31 levels of Zr VI. 
\end{flushleft}
{\small
\begin{tabular}{rllrrrrrrrrrr} \hline
 & & & & & & & &  \\
Index  & Configuration         & Level               & NIST     & GRASP2a & GRASP2b & GRASP3  &  FAC1	& FAC2    &  FAC3     \\
& & & & & & & &   \\ \hline  
& & & & & & & &   \\
  1  & 4s$^2$4p$^5$	       &  $^2$P$^o_{3/2}$    & 0.00000  & 0.00000 & 0.00000 & 0.00000 & 0.00000 & 0.00000 &  0.00000  \\
  2  & 4s$^2$4p$^5$	       &  $^2$P$^o_{1/2}$    & 0.14218  & 0.13790 & 0.13719 & 0.13852 & 0.14422 & 0.14027 &  0.14058  \\
  3  & 4s4p$^6$ 	       &  $^2$S$  _{1/2}$    & 1.74572  & 1.72609 & 1.85015 & 1.84426 & 1.86689 & 1.89504 &  1.89028  \\
  4  & 4s$^2$4p$^4$($^3$P)4d   &  $^4$D$  _{5/2}$    & 2.26849  & 2.22219 & 2.35172 & 2.32398 & 2.38141 & 2.39190 &  2.38799  \\
  5  & 4s$^2$4p$^4$($^3$P)4d   &  $^4$D$  _{7/2}$    & 2.27383  & 2.22329 & 2.35284 & 2.32581 & 2.38421 & 2.39499 &  2.39083  \\
  6  & 4s$^2$4p$^4$($^3$P)4d   &  $^4$D$  _{3/2}$    & 2.27834  & 2.23379 & 2.36336 & 2.33532 & 2.39269 & 2.40310 &  2.39940  \\
  7  & 4s$^2$4p$^4$($^3$P)4d   &  $^4$D$  _{1/2}$    & 2.30038  & 2.25129 & 2.38079 & 2.35322 & 2.41187 & 2.42263 &  2.41897  \\
  8  & 4s$^2$4p$^4$($^3$P)4d   &  $^4$F$  _{9/2}$    & 2.37892  & 2.34759 & 2.47839 & 2.44657 & 2.49998 & 2.51192 &  2.50911  \\
  9  & 4s$^2$4p$^4$($^3$P)4d   &  $^4$F$  _{7/2}$    & 2.42527  & 2.39529 & 2.52620 & 2.49375 & 2.54604 & 2.55851 &  2.55634  \\
 10  & 4s$^2$4p$^4$($^1$D)4d   &  $^2$P$  _{1/2}$    & 2.42652  & 2.42659 & 2.55677 & 2.53249 & 2.57469 & 2.58911 &  2.58904  \\
 11  & 4s$^2$4p$^4$($^3$P)4d   &  $^4$F$  _{5/2}$    & 2.47293  & 2.44719 & 2.56717 & 2.52571 & 2.59208 & 2.60078 &  2.59853  \\
 12  & 4s$^2$4p$^4$($^3$P)4d   &  $^4$F$  _{3/2}$    & 2.47222  & 2.43629 & 2.57785 & 2.53730 & 2.60024 & 2.60221 &  2.60070  \\
 13  & 4s$^2$4p$^4$($^3$P)4d   &  $^4$P$  _{1/2}$    & 2.47949  & 2.47339 & 2.60489 & 2.57427 & 2.61720 & 2.63754 &  2.63878  \\
 14  & 4s$^2$4p$^4$($^3$P)4d   &  $^4$P$  _{3/2}$    & 2.48624  & 2.47679 & 2.60774 & 2.57428 & 2.62344 & 2.63850 &  2.63901  \\
 15  & 4s$^2$4p$^4$($^1$D)4d   &  $^2$D$  _{3/2}$    & 2.50294  & 2.49999 & 2.63063 & 2.59450 & 2.64605 & 2.65788 &  2.65815  \\
 16  & 4s$^2$4p$^4$($^3$P)4d   &  $^2$F$  _{7/2}$    & 2.52597  & 2.50679 & 2.63844 & 2.60400 & 2.65367 & 2.66757 &  2.66690  \\
 17  & 4s$^2$4p$^4$($^3$P)4d   &  $^4$P$  _{5/2}$    & 2.54009  & 2.53399 & 2.66563 & 2.62709 & 2.67898 & 2.69048 &  2.69186  \\
 18  & 4s$^2$4p$^4$($^1$D)4d   &  $^2$P$  _{3/2}$    & 2.54663  & 2.54399 & 2.67479 & 2.64152 & 2.69248 & 2.70698 &  2.70798  \\
 19  & 4s$^2$4p$^4$($^1$D)4d   &  $^2$D$  _{5/2}$    & 2.57992  & 2.57649 & 2.70785 & 2.67242 & 2.72290 & 2.73680 &  2.73795  \\
 20  & 4s$^2$4p$^4$($^1$D)4d   &  $^2$G$  _{7/2}$    & 2.60324  & 2.59749 & 2.73076 & 2.69307 & 2.74738 & 2.75628 &  2.75561  \\
 21  & 4s$^2$4p$^4$($^1$D)4d   &  $^2$G$  _{9/2}$    & 2.60990  & 2.59989 & 2.72800 & 2.69048 & 2.74599 & 2.75289 &  2.75131  \\
 22  & 4s$^2$4p$^4$($^3$P)4d   &  $^2$F$  _{5/2}$    & 2.61664  & 2.61839 & 2.75028 & 2.71456 & 2.76475 & 2.77958 &  2.78065  \\
 23  & 4s$^2$4p$^4$($^1$D)4d   &  $^2$F$  _{5/2}$    & 2.73025  & 2.76328 & 2.89566 & 2.85774 & 2.89540 & 2.91460 &  2.91797  \\
 24  & 4s$^2$4p$^4$($^1$D)4d   &  $^2$F$  _{7/2}$    & 2.76587  & 2.79618 & 2.92883 & 2.89147 & 2.92958 & 2.94951 &  2.95304  \\
 25  & 4s$^2$4p$^4$($^1$S)4d   &  $^2$D$  _{3/2}$    & 2.91002  & 2.98018 & 3.11146 & 2.99315 & 3.10213 & 3.04750 &  3.05187  \\
 26  & 4s$^2$4p$^4$($^1$S)4d   &  $^2$D$  _{5/2}$    & 2.96565  & 3.02848 & 3.15949 & 3.03607 & 3.15469 & 3.09217 &  3.09649  \\
 27  & 4s$^2$4p$^4$($^3$P)4d   &  $^2$P$  _{3/2}$    & 3.09541  & 3.24168 & 3.38303 & 3.21302 & 3.35248 & 3.29129 &  3.29429  \\
 28  & 4s$^2$4p$^4$($^1$D)4d   &  $^2$S$  _{1/2}$    & 3.04998  & 3.26788 & 3.40688 & 3.24837 & 3.36795 & 3.32439 &  3.32751  \\
 29  & 4s$^2$4p$^4$($^3$P)4d   &  $^2$P$  _{1/2}$    & 3.15617  & 3.31168 & 3.46308 & 3.33776 & 3.42794 & 3.39881 &  3.40174  \\
 30  & 4s$^2$4p$^4$($^3$P)4d   &  $^2$D$  _{5/2}$    & 3.13216  & 3.32448 & 3.45499 & 3.27798 & 3.42019 & 3.35821 &  3.35799  \\
 31  & 4s$^2$4p$^4$($^3$P)4d   &  $^2$D$  _{3/2}$    & 3.26390  & 3.43938 & 3.58231 & 3.39676 & 3.55405 & 3.48125 &  3.46307  \\
& & & & & & & & \\ \hline            								                	 
\end{tabular}   								   					       
			      							   					       
\begin{flushleft}													       
{\small
NIST: http://physics.nist.gov/PhysRefData/ASD/levels\_form.html \\
GRASP2a: Singh {\em et al} \cite{sam} \\ 
GRASP2b: present calculations from the {\sc grasp} code with 470 levels\\
GRASP3: present calculations from the {\sc grasp} code with 3990 levels\\
FAC1: present calculations from the {\sc fac} code with 470 \\
FAC2: present calculations from the {\sc fac} code with 3990 \\
FAC3: present calculations from the {\sc fac} code with 12,137 levels \\																       
}															       
\end{flushleft} 

\newpage
\clearpage
\begin{flushleft}
Table 5. Energies (Ryd) for the lowest 31 levels of Nb VII. 
\end{flushleft}
{\small
\begin{tabular}{rllrrrrrrrrrr} \hline
 & & & & & & & &  \\
Index  & Configuration         & Level               & NIST     & GRASP2a & GRASP2b & GRASP3  &  FAC1	& FAC2    &  FAC3     \\
& & & & & & & &   \\ \hline  
& & & & & & & &   \\
  1  & 4s$^2$4p$^5$	       &  $^2$P$^o_{3/2}$    & 0.00000  & 0.00000 & 0.00000 & 0.00000 & 0.00000 & 0.00000 &  0.00000  \\
  2  & 4s$^2$4p$^5$	       &  $^2$P$^o_{1/2}$    & 0.17488  & 0.17060 & 0.16977 & 0.17094 & 0.17670 & 0.17247 &  0.17287  \\
  3  & 4s4p$^6$ 	       &  $^2$S$  _{1/2}$    & 1.93645  & 1.92489 & 2.06099 & 2.04997 & 2.07253 & 2.10509 &  2.10205  \\
  4  & 4s$^2$4p$^4$($^3$P)4d   &  $^4$D$  _{5/2}$    & 2.55304  & 2.51099 & 2.65085 & 2.60419 & 2.67207 & 2.67789 &  2.67506  \\
  5  & 4s$^2$4p$^4$($^3$P)4d   &  $^4$D$  _{7/2}$    &          & 2.51449 & 2.65439 & 2.60858 & 2.67725 & 2.68346 &  2.68036  \\
  6  & 4s$^2$4p$^4$($^3$P)4d   &  $^4$D$  _{3/2}$    & 2.54232  & 2.52429 & 2.66410 & 2.61698 & 2.68483 & 2.69044 &  2.68786  \\
  7  & 4s$^2$4p$^4$($^3$P)4d   &  $^4$D$  _{1/2}$    &          & 2.54649 & 2.68634 & 2.63977 & 2.70860 & 2.71469 &  2.71220  \\
  8  & 4s$^2$4p$^4$($^3$P)4d   &  $^4$F$  _{9/2}$    &          & 2.64749 & 2.78813 & 2.73659 & 2.80212 & 2.80786 &  2.80566  \\
  9  & 4s$^2$4p$^4$($^3$P)4d   &  $^4$F$  _{7/2}$    &          & 2.70258 & 2.84334 & 2.79085 & 2.85515 & 2.86133 &  2.85990  \\
 10  & 4s$^2$4p$^4$($^1$D)4d   &  $^2$P$  _{1/2}$    & 2.72076  & 2.73708 & 2.87723 & 2.82506 & 2.88891 & 2.89561 &  2.89618  \\
 11  & 4s$^2$4p$^4$($^3$P)4d   &  $^4$F$  _{5/2}$    & 2.77699  & 2.76398 & 2.89644 & 2.84123 & 2.91376 & 2.91545 &  2.91394  \\
 12  & 4s$^2$4p$^4$($^3$P)4d   &  $^4$F$  _{3/2}$    & 2.76368  & 2.75578 & 2.90454 & 2.84059 & 2.91775 & 2.91044 &  2.90993  \\
 13  & 4s$^2$4p$^4$($^3$P)4d   &  $^4$P$  _{1/2}$    & 2.77662  & 2.78998 & 2.93113 & 2.87934 & 2.93694 & 2.94914 &  2.95076  \\
 14  & 4s$^2$4p$^4$($^3$P)4d   &  $^4$P$  _{3/2}$    & 2.79056  & 2.79538 & 2.93614 & 2.88158 & 2.94594 & 2.95299 &  2.95383  \\
 15  & 4s$^2$4p$^4$($^1$D)4d   &  $^2$D$  _{3/2}$    & 2.80736  & 2.82038 & 2.96094 & 2.90529 & 2.97054 & 2.97584 &  2.97676  \\
 16  & 4s$^2$4p$^4$($^3$P)4d   &  $^2$F$  _{7/2}$    &          & 2.82938 & 2.97053 & 2.91550 & 2.98041 & 2.98627 &  2.98598  \\
 17  & 4s$^2$4p$^4$($^3$P)4d   &  $^4$P$  _{5/2}$    & 2.84901  & 2.86288 & 3.00406 & 2.94241 & 3.01047 & 3.01205 &  3.01387  \\
 18  & 4s$^2$4p$^4$($^1$D)4d   &  $^2$P$  _{3/2}$    & 2.86238  & 2.87828 & 3.01898 & 2.96455 & 3.03036 & 3.03694 &  3.03854  \\
 19  & 4s$^2$4p$^4$($^1$D)4d   &  $^2$D$  _{5/2}$    & 2.89898  & 2.91348 & 3.05449 & 2.99912 & 3.06353 & 3.07048 &  3.07224  \\
 20  & 4s$^2$4p$^4$($^1$D)4d   &  $^2$G$  _{7/2}$    &          & 2.93298 & 3.07359 & 3.01492 & 3.08355 & 3.08488 &  3.08490  \\
 21  & 4s$^2$4p$^4$($^1$D)4d   &  $^2$G$  _{9/2}$    &          & 2.93218 & 3.07265 & 3.01464 & 3.08432 & 3.08417 &  3.08334  \\
 22  & 4s$^2$4p$^4$($^3$P)4d   &  $^2$F$  _{5/2}$    & 2.93916  & 2.96018 & 3.10151 & 3.04360 & 3.10941 & 3.11496 &  3.11651  \\
 23  & 4s$^2$4p$^4$($^1$D)4d   &  $^2$F$  _{5/2}$    & 3.05794  & 3.11078 & 3.25236 & 3.19222 & 3.24718 & 3.25650 &  3.26021  \\
 24  & 4s$^2$4p$^4$($^1$D)4d   &  $^2$F$  _{7/2}$    &          & 3.15278 & 3.29456 & 3.23499 & 3.29042 & 3.30049 &  3.30435  \\
 25  & 4s$^2$4p$^4$($^1$S)4d   &  $^2$D$  _{3/2}$    & 3.25314  & 3.34238 & 3.48307 & 3.34059 & 3.47023 & 3.40275 &  3.40749  \\
 26  & 4s$^2$4p$^4$($^1$S)4d   &  $^2$D$  _{5/2}$    & 3.32687  & 3.40598 & 3.54658 & 3.39929 & 3.53861 & 3.46374 &  3.46846  \\
 27  & 4s$^2$4p$^4$($^1$D)4d   &  $^2$S$  _{1/2}$    & 3.40216  & 3.64738 & 3.80087 & 3.63579 & 3.75984 & 3.71226 &  3.71659  \\
 28  & 4s$^2$4p$^4$($^3$P)4d   &  $^2$P$  _{3/2}$    & 3.47197  & 3.65598 & 3.79497 & 3.59877 & 3.76239 & 3.67961 &  3.68365  \\
 29  & 4s$^2$4p$^4$($^3$P)4d   &  $^2$P$  _{1/2}$    & 3.54432  & 3.72368 & 3.87262 & 3.72157 & 3.84506 & 3.78908 &  3.79379  \\
 30  & 4s$^2$4p$^4$($^3$P)4d   &  $^2$D$  _{5/2}$    & 3.51492  & 3.73168 & 3.87813 & 3.67645 & 3.83604 & 3.75806 &  3.75930  \\
 31  & 4s$^2$4p$^4$($^3$P)4d   &  $^2$D$  _{3/2}$    & 3.67477  & 3.87958 & 4.02827 & 3.82167 & 3.99813 & 3.90759 &  3.90984  \\
& & & & & & & & \\ \hline            								                	 
\end{tabular}   								   					       
			      							   					       
\begin{flushleft}													       
{\small
NIST: http://physics.nist.gov/PhysRefData/ASD/levels\_form.html \\
GRASP2a: Singh {\em et al} \cite{sam} \\ 
GRASP2b: present calculations from the {\sc grasp} code with 470 levels\\
GRASP3: present calculations from the {\sc grasp} code with 3990 levels\\
FAC1: present calculations from the {\sc fac} code with 470 levels  \\  
FAC2: present calculations from the {\sc fac} code with 3990 levels \\ 													
FAC3: present calculations from the {\sc fac} code with 12,137 levels \\																       
}															       
\end{flushleft} 

\newpage
\clearpage

\begin{flushleft}
Table 6. Energies (Ryd) for the lowest 31 levels of Mo VIII. 
\end{flushleft}
{\small
\begin{tabular}{rllrrrrrrrrrr} \hline
 & & & & & & & &  \\
Index  & Configuration         & Level               & NIST     & GRASP2a & GRASP2b & GRASP3  &  FAC1	& FAC2    & FAC3     \\
& & & & & & & &   \\ \hline  
& & & & & & & &   \\
  1  & 4s$^2$4p$^5$	       &  $^2$P$^o_{3/2}$    & 0.00000  & 0.00000 & 0.00000 & 0.00000 & 0.00000 & 0.00000 & 0.00000  \\
  2  & 4s$^2$4p$^5$	       &  $^2$P$^o_{1/2}$    & 0.21209  & 0.20770 & 0.20676 & 0.20776 & 0.21363 & 0.20913 & 0.20962  \\
  3  & 4s4p$^6$ 	       &  $^2$S$  _{1/2}$    & 2.13082  & 2.12679 & 2.27306 & 2.25774 & 2.28111 & 2.31642 & 2.31460  \\
  4  & 4s$^2$4p$^4$($^3$P)4d   &  $^4$D$  _{5/2}$    & 2.81307  & 2.79438 & 2.94319 & 2.88062 & 2.95802 & 2.95864 & 2.95659  \\
  5  & 4s$^2$4p$^4$($^3$P)4d   &  $^4$D$  _{7/2}$    &          & 2.80088 & 2.94980 & 2.88823 & 2.96622 & 2.96738 & 2.96505  \\
  6  & 4s$^2$4p$^4$($^3$P)4d   &  $^4$D$  _{3/2}$    & 2.82436  & 2.80908 & 2.95788 & 2.89461 & 2.97206 & 2.97229 & 2.97051  \\
  7  & 4s$^2$4p$^4$($^3$P)4d   &  $^4$D$  _{1/2}$    &          & 2.83678 & 2.98561 & 2.92291 & 3.00109 & 3.00186 & 3.00021  \\
  8  & 4s$^2$4p$^4$($^3$P)4d   &  $^4$F$  _{9/2}$    &          & 2.94108 & 3.09038 & 3.02231 & 3.09862 & 3.09795 & 3.09621  \\
  9  & 4s$^2$4p$^4$($^3$P)4d   &  $^4$F$  _{7/2}$    &          & 3.00278 & 3.15221 & 3.08273 & 3.15778 & 3.15727 & 3.15638  \\
 10  & 4s$^2$4p$^4$($^1$D)4d   &  $^2$P$  _{1/2}$    & 3.01450  & 3.04338 & 3.19233 & 3.12243 & 3.19970 & 3.19842 & 3.19928  \\
 12  & 4s$^2$4p$^4$($^3$P)4d   &  $^4$F$  _{3/2}$    & 3.05879  & 3.07138 & 3.22305 & 3.13829 & 3.22889 & 3.21173 & 3.21195  \\
 11  & 4s$^2$4p$^4$($^3$P)4d   &  $^4$F$  _{5/2}$    & 3.07954  & 3.10018 & 3.22072 & 3.14797 & 3.23226 & 3.22651 & 3.22557  \\
 13  & 4s$^2$4p$^4$($^3$P)4d   &  $^4$P$  _{1/2}$    & 3.07039  & 3.07388 & 3.24975 & 3.17990 & 3.25085 & 3.25444 & 3.25614  \\
 14  & 4s$^2$4p$^4$($^3$P)4d   &  $^4$P$  _{3/2}$    & 3.09398  & 3.10968 & 3.25906 & 3.18700 & 3.26494 & 3.26378 & 3.26468  \\
 15  & 4s$^2$4p$^4$($^1$D)4d   &  $^2$D$  _{3/2}$    & 3.11072  & 3.13738 & 3.28654 & 3.21432 & 3.29200 & 3.29015 & 3.29142  \\
 16  & 4s$^2$4p$^4$($^3$P)4d   &  $^2$F$  _{7/2}$    &          & 3.14868 & 3.29824 & 3.22612 & 3.30440 & 3.30240 & 3.30236  \\
 17  & 4s$^2$4p$^4$($^3$P)4d   &  $^4$P$  _{5/2}$    & 3.15494  & 3.18568 & 3.33528 & 3.25358 & 3.33650 & 3.32775 & 3.32973  \\
 18  & 4s$^2$4p$^4$($^1$D)4d   &  $^2$P$  _{3/2}$    & 3.17879  & 3.20998 & 3.35937 & 3.28684 & 3.36623 & 3.36437 & 3.36621  \\
 19  & 4s$^2$4p$^4$($^1$D)4d   &  $^2$D$  _{5/2}$    & 3.21812  & 3.24758 & 3.39715 & 3.32473 & 3.40180 & 3.40120 & 3.40326  \\
 20  & 4s$^2$4p$^4$($^1$D)4d   &  $^2$G$  _{7/2}$    &          & 3.26258 & 3.41187 & 3.33549 & 3.41688 & 3.41039 & 3.41091  \\
 21  & 4s$^2$4p$^4$($^1$D)4d   &  $^2$G$  _{9/2}$    &          & 3.26488 & 3.41404 & 3.33888 & 3.42098 & 3.41371 & 3.41344  \\
 22  & 4s$^2$4p$^4$($^3$P)4d   &  $^2$F$  _{5/2}$    & 3.26061  & 3.29768 & 3.44736 & 3.37036 & 3.45040 & 3.44632 & 3.44811  \\
 23  & 4s$^2$4p$^4$($^1$D)4d   &  $^2$F$  _{5/2}$    & 3.38391  & 3.45298 & 3.60277 & 3.52373 & 3.59423 & 3.59342 & 3.59720  \\
 24  & 4s$^2$4p$^4$($^1$D)4d   &  $^2$F$  _{7/2}$    &          & 3.50498 & 3.65493 & 3.57645 & 3.64736 & 3.64733 & 3.65125  \\
 25  & 4s$^2$4p$^4$($^1$S)4d   &  $^2$D$  _{3/2}$    & 3.59536  & 3.70008 & 3.84920 & 3.68624 & 3.83405 & 3.75434 & 3.75919  \\
 26  & 4s$^2$4p$^4$($^1$S)4d   &  $^2$D$  _{5/2}$    & 3.68975  & 3.78268 & 3.93173 & 3.76445 & 3.92182 & 3.83531 & 3.84018  \\
 27  & 4s$^2$4p$^4$($^1$D)4d   &  $^2$S$  _{1/2}$    & 3.74997  & 4.02958 & 4.18120 & 4.01094 & 4.14037 & 4.08705 & 4.09204  \\
 28  & 4s$^2$4p$^4$($^3$P)4d   &  $^2$P$  _{3/2}$    & 3.84153  & 4.04118 & 4.19501 & 3.97538 & 4.16110 & 4.05783 & 4.06231  \\
 29  & 4s$^2$4p$^4$($^3$P)4d   &  $^2$P$  _{1/2}$    & 3.92728  & 4.12238 & 4.28906 & 4.10027 & 4.25591 & 4.17334 & 4.17883  \\
 30  & 4s$^2$4p$^4$($^3$P)4d   &  $^2$D$  _{5/2}$    & 3.88909  & 4.13578 & 4.27715 & 4.06291 & 4.23931 & 4.14535 & 4.14748  \\
 31  & 4s$^2$4p$^4$($^3$P)4d   &  $^2$D$  _{3/2}$    & 4.08135  & 4.31058 & 4.46526 & 4.23847 & 4.43382 & 4.32532 & 4.32852  \\
& & & & & & & & \\ \hline            								                	 
\end{tabular}   								   					       
			      							   					       
\begin{flushleft}													       
{\small
NIST: http://physics.nist.gov/PhysRefData/ASD/levels\_form.html \\
GRASP2a: Singh {\em et al} \cite{sam} \\ 
GRASP2b: present calculations from the {\sc grasp} code with 470 levels\\
GRASP3: present calculations from the {\sc grasp} code with 3990 levels\\
FAC1: present calculations from the {\sc fac} code with 470 levels  \\  
FAC2: present calculations from the {\sc fac} code with 3990 levels \\ 													
FAC3: present calculations from the {\sc fac} code with 12,137 levels \\																       
}															       
\end{flushleft} 

\newpage
\clearpage

\begin{flushleft}
Table 7. Comparison of radiative rates (A- values, s$^{-1}$) for E1 transitions among levels of  the ground state 4s$^2$4p$^5$ $^2$P$^o_{3/2,1/2}$ and 4s4p$^6$ and 4s$^2$4p$^4$4d configurations of Sr IV -- see Table 2 for level definitions.  $a{\pm}b \equiv a{\times}$10$^{{\pm}b}$.
\end{flushleft}
{\small
\begin{tabular}{rrllllllllllll} \hline
 \multicolumn{2}{c}{Transition} & GRASP2a  & GRASP2b   & \multicolumn{2}{c}{GRASP3}  & FAC2   & FAC3\\ \hline
   I &     J &     A       &	 A	 &     A       &   f	     & A	  & A	      &  Ratio  \\
\hline
   1 &     3 &    1.55+08  &  3.6342+08  &  3.5713+08  &  1.0660-02  & 4.077+08   & 4.074+08  & 2.2-1  \\
   1 &     5 &    5.27+06  &  5.9143+06  &  1.0231+07  &  6.1860-04  & 1.199+07   & 1.181+07  & 9.4-1  \\
   1 &     6 &    2.57+06  &  2.4018+06  &  2.5331+06  &  1.0117-04  & 4.352+06   & 4.138+06  & 7.8-1  \\
   1 &     7 &    1.02+06  &  8.0619+05  &  4.1581+05  &  8.2090-06  & 1.605+06   & 1.383+06  & 4.8-1  \\
   1 &    10 &    5.83+06  &  4.4562+06  &  3.3635+05  &  5.6538-06  & 1.857+06   & 6.461+05  & 1.9-0  \\
   1 &    11 &    5.38+07  &  4.6873+07  &  8.1160+07  &  4.1389-03  & 1.073+08   & 1.047+08  & 9.1-1  \\
   1 &    12 &    5.38+06  &  6.5981+06  &  8.0781+06  &  2.7229-04  & 1.591+06   & 3.035+06  & 9.4-1  \\
   1 &    13 &    1.60+08  &  1.4338+08  &  2.5878+08  &  4.2103-03  & 2.979+08   & 3.068+08  & 7.9-1  \\
   1 &    14 &    2.19+08  &  1.6215+08  &  1.7587+08  &  5.7215-03  & 2.437+08   & 2.049+08  & 8.7-1  \\
   1 &    15 &    2.01+08  &  1.4387+08  &  2.1917+08  &  7.0193-03  & 1.689+08   & 1.644+08  & 9.4-1  \\
   1 &    17 &    7.00+07  &  4.9727+07  &  7.6048+07  &  3.5911-03  & 6.909+07   & 5.897+07  & 8.1-1  \\
   1 &    18 &    1.00+08  &  7.4910+07  &  6.0688+04  &  1.8950-06  & 1.481+05   & 8.054+05  & 4.2-0  \\
   1 &    19 &    3.43+08  &  2.2873+08  &  2.3311+08  &  1.0667-02  & 1.659+08   & 1.552+08  & 9.1-1  \\
   1 &    20 &    1.57+08  &  1.2299+08  &  1.6340+08  &  7.2302-03  & 1.331+08   & 1.117+08  & 9.7-1  \\
   1 &    23 &    1.59+08  &  1.3933+08  &  2.3389+08  &  9.1410-03  & 3.224+08   & 2.939+08  & 8.1-1  \\
   1 &    25 &    6.12+08  &  4.7105+08  &  1.9286+09  &  4.5624-02  & 1.336+09   & 2.203+09  & 9.7-1  \\
   1 &    26 &    3.76+08  &  2.5017+08  &  1.5860+09  &  5.5298-02  & 1.082+09   & 1.729+09  & 1.0-0  \\
   1 &    27 &    8.91+10  &  6.9511+10  &  5.7718+10  &  1.2294-00  & 6.832+10   & 6.554+10  & 9.1-1  \\   
   1 &    28 &    5.15+10  &  3.8773+10  &  3.0543+10  &  3.1811-01  & 4.191+10   & 4.402+10  & 9.0-1  \\
   1 &    29 &    1.08+11  &  9.1973+10  &  5.7973+10  &  1.7984-00  & 7.551+10   & 7.786+10  & 8.8-1  \\   
   1 &    30 &    5.39+10  &  5.4270+10  &  3.4522+10  &  3.3963-01  & 3.371+10   & 2.042+10  & 8.1-1  \\
   1 &    31 &    4.98+09  &  4.5917+09  &  2.5073+09  &  4.9008-02  & 2.771+09   & 2.378+09  & 8.3-1  \\
   2 &     3 &    7.79+07  &  1.7135+08  &  1.7377+08  &  1.1714-02  & 1.989+08   & 1.977+08  & 2.0-1  \\
   2 &     6 &    4.86+04  &  2.3776+04  &  2.9543+05  &  2.6047-05  & 3.689+05   & 4.498+05  & 1.3-0  \\
   2 &     7 &    1.33+06  &  9.7317+05  &  3.4896+05  &  1.5200-05  & 8.247+05   & 6.745+05  & 4.6-1  \\
   2 &    10 &    3.05+07  &  2.5164+07  &  3.7542+06  &  1.3816-04  & 1.106+07   & 8.324+06  & 4.9-1  \\
   2 &    12 &    2.16+07  &  2.2836+07  &  3.1660+07  &  2.3367-03  & 3.526+07   & 3.536+07  & 9.7-1  \\
   2 &    13 &    4.30+06  &  6.4777+06  &  3.0353+06  &  1.0796-04  & 7.175+06   & 7.375+06  & 1.8-1  \\
   2 &    14 &    1.59+07  &  1.4550+07  &  7.7999+05  &  5.5470-05  & 8.083+05   & 6.851+05  & 1.3-0  \\
   2 &    15 &    2.61+08  &  1.8430+08  &  2.5448+08  &  1.7804-02  & 2.484+08   & 2.282+08  & 9.1-1  \\
   2 &    18 &    5.65+07  &  3.7116+07  &  3.0459+07  &  2.0753-03  & 1.150+07   & 1.228+07  & 9.3-1  \\
   2 &    25 &    1.99+09  &  1.5167+09  &  3.5455+08  &  1.8092-02  & 3.779+08   & 4.531+08  & 6.5-1  \\
   2 &    27 &    2.60+09  &  1.7449+09  &  6.5874+08  &  3.0148-02  & 1.305+09   & 7.921+08  & 1.1-0  \\
   2 &    28 &    4.16+10  &  3.3517+10  &  2.7800+10  &  6.2157-01  & 2.895+10   & 2.360+10  & 9.3-1  \\
   2 &    30 &    5.37+10  &  4.5666+10  &  3.7795+10  &  7.9664-01  & 4.680+10   & 3.969+10  & 8.3-1  \\   
   2 &    31 &    1.01+11  &  8.5572+10  &  7.0119+10  &  2.9356-00  & 8.032+10   & 7.831+10  & 8.8-1  \\ 
\hline            								                	 
\end{tabular}   								   					       
			      							   					       
\begin{flushleft}													       
{\small
GRASP2a: Singh {\em et al} \cite{sam} \\ 
GRASP2b: present calculations from the {\sc grasp} code with 470 levels\\
GRASP3: present calculations from the {\sc grasp} code with 3990 levels\\
FAC2: present calculations from the {\sc fax} code with 3990 levels\\
FAC3: present calculations from the {\sc fax} code with 12,137 levels\\
Ratio: ratio of velocity and length forms of A- values \\																       
}															       
\end{flushleft} 

\newpage
\clearpage

\begin{flushleft}
Table 8. Comparison of radiative rates (A- values, s$^{-1}$) for E1 transitions among levels of  the ground state 4s$^2$4p$^5$ $^2$P$^o_{3/2,1/2}$ and 4s4p$^6$ and 4s$^2$4p$^4$4d configurations of Y V -- see Table 3 for level definitions.  $a{\pm}b \equiv a{\times}$10$^{{\pm}b}$.
\end{flushleft}
{\small
\begin{tabular}{rrllllllll} \hline
 & & & & & & & &  \\
 \multicolumn{2}{c}{Transition} & GRASP2a  & GRASP2b & \multicolumn{3}{c}{GRASP3} \\ \hline
   I &     J &     A       &        A	    &       A	    &	f	 & Ratio   \\
\hline  
     1 &     3 &   3.41+08 &     7.4178+08 &    6.4885+08  &  1.5026-02  & 4.0-1 \\
     1 &     4 &   8.49+06 &     1.0049+07 &    1.4695+07  &  6.6085-04  & 9.5-1 \\
     1 &     6 &   5.17+06 &     5.0984+06 &    5.7124+06  &  1.6963-04  & 9.3-1 \\
     1 &     7 &   2.69+06 &     2.3091+06 &    1.9727+06  &  2.8901-05  & 8.8-1 \\
     1 &    10 &   1.10+07 &     9.1719+06 &    2.4808+06  &  3.1312-05  & 1.4-0 \\
     1 &    11 &   9.04+07 &     8.3561+07 &    1.3428+08  &  5.0927-03  & 9.5-1 \\
     1 &    12 &   2.44+06 &     5.2851+06 &    2.4683+06  &  6.1990-05  & 1.0-0 \\
     1 &    13 &   2.90+08 &     2.7110+08 &    3.9343+08  &  4.7899-03  & 8.1-1 \\
     1 &    14 &   3.73+08 &     2.9292+08 &    4.1741+08  &  1.0164-02  & 9.5-1 \\
     1 &    15 &   2.77+08 &     2.0450+08 &    2.3712+08  &  5.6861-03  & 9.3-1 \\
     1 &    17 &   1.10+08 &     7.8261+07 &    1.2373+08  &  4.3518-03  & 8.4-1 \\
     1 &    18 &   8.38+07 &     6.4620+07 &    6.7013+06  &  1.5578-04  & 1.2-0 \\
     1 &    19 &   3.66+08 &     2.4426+08 &    2.4821+08  &  8.4570-03  & 9.6-1 \\
     1 &    22 &   1.49+08 &     1.1934+08 &    2.2968+08  &  7.5744-03  & 1.0-0 \\
     1 &    23 &   2.34+08 &     2.1489+08 &    3.3501+08  &  9.8810-03  & 8.3-1 \\
     1 &    25 &   1.06+09 &     8.6167+08 &    3.1288+09  &  5.5895-02  & 9.8-1 \\
     1 &    26 &   5.28+07 &     1.8626+07 &    1.4890+09  &  3.8992-02  & 1.1-0 \\
     1 &    27 &   1.27+11 &     1.0138+11 &    1.0669+11  &  1.6837-00  & 9.2-1 \\   
     1 &    28 &   1.07+11 &     8.2858+10 &    6.8148+10  &  5.2522-01  & 9.0-1 \\
     1 &    29 &   1.55+11 &     1.3485+11 &    1.2220+11  &  2.7893-00  & 9.0-1 \\   
     1 &    30 &   3.70+10 &     4.5026+10 &    5.8584+10  &  4.2274-01  & 8.1-1 \\
     1 &    31 &   6.68+09 &     6.2932+09 &    4.2526+09  &  6.0599-02  & 8.7-1 \\
     2 &     3 &   1.59+08 &     3.3294+08 &    3.0194+08  &  1.6071-02  & 3.7-1 \\
     2 &     6 &   1.47+05 &     1.0055+04 &    7.9784+04  &  5.2924-06  & 1.1-0 \\
     2 &     7 &   3.11+06 &     2.4823+06 &    1.7391+06  &  5.6874-05  & 1.0-0 \\
     2 &    10 &   5.58+07 &     4.8736+07 &    2.7075+07  &  7.5666-04  & 1.0-0 \\
     2 &    12 &   2.89+07 &     3.3965+07 &    4.0470+07  &  2.2499-03  & 1.0-0 \\
     2 &    13 &   1.30+07 &     1.7692+07 &    1.1362+07  &  3.0570-04  & 5.0-1 \\
     2 &    14 &   2.24+07 &     2.1717+07 &    2.7072+06  &  1.4570-04  & 1.5-0 \\
     2 &    15 &   3.20+08 &     2.2841+08 &    3.2991+08  &  1.7471-02  & 9.7-1 \\
     2 &    18 &   2.86+07 &     1.6216+07 &    1.7387+07  &  8.9120-04  & 8.7-1 \\
     2 &    25 &   1.92+09 &     1.4732+09 &    5.6428+08  &  2.1956-02  & 7.0-1 \\
     2 &    27 &   3.32+09 &     2.2497+09 &    2.3159+09  &  7.9184-02  & 9.9-1 \\
     2 &    28 &   2.85+10 &     2.6514+10 &    4.1111+10  &  6.8582-01  & 9.5-1 \\
     2 &    30 &   1.03+11 &     8.4428+10 &    3.9796+10  &  6.1996-01  & 8.6-1 \\   
     2 &    31 &   1.46+11 &     1.2577+11 &    1.2065+11  &  3.7097-00  & 9.0-1 \\ 
\hline            								                	 
\end{tabular}   								   					       
			      							   					       
\begin{flushleft}													       
{\small
GRASP2a: Singh {\em et al} \cite{sam} \\ 
GRASP2b: present calculations from the {\sc grasp} code with 470 levels\\
GRASP3: present calculations from the {\sc grasp} code with 3990 levels\\
Ratio: ratio of velocity and length forms of A -values \\																       
}															       
\end{flushleft} 

\newpage
\clearpage

\begin{flushleft}
Table 9. Comparison of radiative rates (A- values, s$^{-1}$) for E1 transitions among levels of  the ground state 4s$^2$4p$^5$ $^2$P$^o_{3/2,1/2}$ and 4s4p$^6$ and 4s$^2$4p$^4$4d configurations of Zr VI -- see Table 4 for level definitions.  $a{\pm}b \equiv a{\times}$10$^{{\pm}b}$.
\end{flushleft}
{\small
\begin{tabular}{rrllllllll} \hline
 & & & & & & & &  \\
 \multicolumn{2}{c}{Transition} & GRASP2a  & GRASP2b & \multicolumn{3}{c}{GRASP3} \\ \hline
   I &     J &   A     &     A      &     A     &   f       & Ratio  \\
\hline  
   1  &  3 &  5.30+08  & 1.2340+09  & 1.0518+09 & 1.9249-02 & 4.4-1 \\
   1  &  4 &  1.17+07  & 1.5391+07  & 2.0308+07 & 7.0217-04 & 9.4-1 \\
   1  &  6 &  7.76+06  & 9.2931+06  & 1.1017+07 & 2.5149-04 & 9.8-1 \\
   1  &  7 &  4.75+06  & 5.3458+06  & 5.6982+06 & 6.4052-05 & 9.8-1 \\
   1  & 10 &  1.28+07  & 1.4776+07  & 2.1200+08 & 6.1728-03 & 9.6-1 \\
   1  & 11 &  1.41+08  & 1.4199+08  & 1.0789+07 & 1.0527-04 & 1.4-0 \\
   1  & 12 &  3.35+05  & 1.0270+06  & 5.1806+05 & 1.0018-05 & 9.2-1 \\
   1  & 13 &  4.67+08  & 4.7004+08  & 5.9759+08 & 5.6133-03 & 8.0-1 \\
   1  & 14 &  5.50+08  & 4.8841+08  & 7.0038+08 & 1.3157-02 & 9.8-1 \\
   1  & 15 &  3.37+08  & 2.5187+08  & 2.5390+08 & 4.6958-03 & 9.1-1 \\
   1  & 17 &  1.60+08  & 1.1567+08  & 1.5603+08 & 4.2219-03 & 8.3-1 \\
   1  & 18 &  6.97+07  & 6.2901+07  & 1.7608+07 & 3.1416-04 & 1.3-0 \\
   1  & 19 &  3.82+08  & 2.5563+08  & 2.6359+08 & 6.8923-03 & 9.8-1 \\
   1  & 22 &  1.96+08  & 1.4639+08  & 3.0306+08 & 7.6802-03 & 1.0-0 \\
   1  & 23 &  3.24+08  & 3.1381+08  & 3.9608+08 & 9.0568-03 & 8.2-1 \\
   1  & 25 &  1.68+09  & 1.4350+09  & 4.0018+09 & 5.5609-02 & 9.9-1 \\
   1  & 26 &  3.16+02  & 2.1868+07  & 2.1104+09 & 4.2754-02 & 1.1-0 \\
   1  & 27 &  1.56+11  & 1.3479+11  & 1.2527+11 & 1.5107-00 & 9.3-1 \\   
   1  & 28 &  1.63+11  & 1.5331+11  & 9.7116+10 & 5.7290-01 & 9.0-1 \\
   1  & 29 &  1.91+11  & 1.0623+10  & 6.3978+10 & 3.5747-01 & 8.0-1 \\   
   1  & 30 &  1.29+10  & 1.7896+11  & 1.6154+11 & 2.8075-00 & 9.1-1 \\
   1  & 31 &  7.17+09  & 7.6240+09  & 7.1045+09 & 7.6658-02 & 8.8-1 \\
   2  &  3 &  2.40+08  & 5.3588+08  & 4.7253+08 & 2.0219-02 & 4.1-1 \\
   2  &  6 &  3.21+05  & 1.0080+03  & 1.2890-00 & 6.6505-11 & 1.2+3 \\
   2  &  7 &  4.55+06  & 4.9856+06  & 4.8266+06 & 1.2251-04 & 1.1-0 \\
   2  & 10 &  6.32+07  & 7.2178+07  & 6.0598+07 & 1.3238-03 & 1.1-0 \\
   2  & 12 &  3.56+07  & 4.7038+07  & 5.1666+07 & 2.2356-03 & 1.1-0 \\
   2  & 13 &  2.37+07  & 3.6522+07  & 2.6970+07 & 5.6593-04 & 6.0-1 \\
   2  & 14 &  1.58+07  & 2.0358+07  & 3.2332+06 & 1.3569-04 & 1.7-0 \\
   2  & 15 &  3.78+08  & 2.7949+08  & 4.0027+08 & 1.6523-02 & 9.9-1 \\
   2  & 18 &  2.06+07  & 8.4120+06  & 1.1180+07 & 4.4433-04 & 7.5-1 \\
   2  & 25 &  2.20+09  & 1.6511+09  & 6.9689+08 & 2.1293-02 & 7.0-1 \\
   2  & 27 &  3.20+09  & 2.4759+09  & 2.5940+09 & 6.8328-02 & 1.0-0 \\
   2  & 28 &  8.16+09  & 3.4364+09  & 3.9101+10 & 5.0334-01 & 9.8-1 \\
   2  & 29 &  1.52+11  & 1.4005+11  & 9.6829+10 & 1.1778-00 & 8.6-1 \\   
   2  & 31 &  1.80+11  & 1.6739+11  & 1.5364+11 & 3.6034-00 & 9.1-1 \\ 
\hline            								                	 
\end{tabular}   								   					       
			      							   					       
\begin{flushleft}													       
{\small
GRASP2a: Singh {\em et al} \cite{sam} \\ 
GRASP2b: present calculations from the {\sc grasp} code with 470 levels\\
GRASP3: present calculations from the {\sc grasp} code with 3990 levels\\

Ratio: ratio of velocity and length forms of A- values \\																       
}															       
\end{flushleft} 

\newpage
\clearpage

\begin{flushleft}
Table 10. Comparison of radiative rates (A- values, s$^{-1}$) for E1 transitions among levels of  the ground state 4s$^2$4p$^5$ $^2$P$^o_{3/2,1/2}$ and 4s4p$^6$ and 4s$^2$4p$^4$4d configurations of Nb VII -- see Table 5 for level definitions.  $a{\pm}b \equiv a{\times}$10$^{{\pm}b}$.
\end{flushleft}
{\small
\begin{tabular}{rrllllllll} \hline
 & & & & & & & &  \\
 \multicolumn{2}{c}{Transition} & GRASP2a  & GRASP2b & \multicolumn{2}{c}{GRASP3} \\ \hline
   I &     J &     A      &      A     &     A     &   f       & Ratio \\
\hline  
     1  &  3  & 8.66+08   & 1.8359+09  & 1.5435+09 & 2.2863-02 & 4.5-1 \\
     1  &  4  & 1.68+07   & 2.1932+07  & 2.6950+07 & 7.4210-04 & 9.3-1 \\
     1  &  6  & 1.34+07   & 1.5424+07  & 1.8777+07 & 3.4134-04 & 1.0-0 \\
     1  &  7  & 1.02+07   & 1.1016+07  & 1.2926+07 & 1.1546-04 & 9.9-1 \\
     1  & 10  & 1.92+07   & 2.0712+07  & 2.0504+07 & 1.5992-04 & 1.4-0 \\
     1  & 11  & 2.29+08   & 2.3196+08  & 3.2927+08 & 7.6169-03 & 9.5-1 \\
     1  & 12  & 7.74+06   & 3.7905+06  & 1.9836+07 & 3.0605-04 & 9.6-1 \\
     1  & 13  & 7.65+08   & 7.6492+08  & 8.9955+08 & 6.7540-03 & 7.9-1 \\
     1  & 14  & 8.57+08   & 7.5394+08  & 1.0257+09 & 1.5378-02 & 9.8-1 \\
     1  & 15  & 3.83+08   & 2.8801+08  & 2.9256+08 & 4.3151-03 & 8.9-1 \\
     1  & 17  & 2.27+08   & 1.6263+08  & 1.9368+08 & 4.1775-03 & 8.1-1 \\
     1  & 18  & 7.40+07   & 6.7028+07  & 3.0557+07 & 4.3286-04 & 1.3-0 \\
     1  & 19  & 4.16+08   & 2.7767+08  & 3.0418+08 & 6.3152-03 & 9.9-1 \\
     1  & 22  & 2.52+08   & 1.9994+08  & 4.1267+08 & 8.3189-03 & 1.0-0 \\
     1  & 23  & 4.50+08   & 4.3792+08  & 4.8703+08 & 8.9250-03 & 8.1-1 \\
     1  & 25  & 2.62+09   & 2.2474+09  & 5.3405+09 & 5.9578-02 & 9.9-1 \\
     1  & 26  & 8.11+07   & 1.6709+08  & 2.9244+09 & 4.7261-02 & 1.0-0 \\
     1  & 27  & 1.98+11   & 2.0079+11  & 1.4062+11 & 6.6216-01 & 8.9-1 \\   
     1  & 28  & 2.18+11   & 1.6914+11  & 1.5224+11 & 1.4635-00 & 9.3-1 \\
     1  & 29  & 2.41+11   & 2.2363+11  & 5.2042+10 & 2.3390-01 & 7.8-1 \\   
     1  & 30  & 1.84+07   & 2.3751+06  & 1.9979+11 & 2.7603-00 & 9.1-1 \\
     1  & 31  & 8.15+09   & 8.5737+09  & 8.3648+09 & 7.1302-02 & 8.8-1 \\
     2  &  3  & 3.77+08   & 7.7571+08  & 6.7475+08 & 2.3791-02 & 4.2-1 \\
     2  &  6  & 8.18+05   & 5.8651+04  & 1.2640+05 & 5.2601-06 & 1.4-0 \\
     2  &  7  & 8.45+06   & 8.7486+06  & 9.8501+06 & 2.0119-04 & 1.2-0 \\
     2  & 10  & 8.66+07   & 9.3367+07  & 9.1692+07 & 1.6205-03 & 1.2-0 \\
     2  & 12  & 4.28+07   & 6.0657+07  & 6.3245+07 & 2.2095-03 & 1.1-0 \\
     2  & 13  & 4.60+07   & 6.5045+07  & 5.0617+07 & 8.5904-04 & 6.3-1 \\
     2  & 14  & 8.79+06   & 1.1772+07  & 1.5539+06 & 5.2657-05 & 2.2-0 \\
     2  & 15  & 4.56+08   & 3.4446+08  & 4.7877+08 & 1.5944-02 & 1.0-0 \\
     2  & 18  & 1.54+07   & 5.5585+06  & 9.2858+06 & 2.9626-04 & 6.6-1 \\
     2  & 25  & 2.61+09   & 2.0128+09  & 9.8756+08 & 2.4475-02 & 7.3-1 \\
     2  & 27  & 3.32+09   & 1.4226+09  & 2.9884+10 & 3.0990-01 & 1.0-0 \\
     2  & 28  & 3.20+08   & 2.4534+09  & 2.7930+09 & 5.9185-02 & 1.0-0 \\
     2  & 29  & 1.99+11   & 1.7491+11  & 1.3456+11 & 1.3287-00 & 8.7-1 \\   
     2  & 31  & 2.28+11   & 2.0973+11  & 1.9008+11 & 3.5510-00 & 9.1-1 \\ 
\hline            								                	 
\end{tabular}   								   					       
			      							   					       
\begin{flushleft}													       
{\small
GRASP2a: Singh {\em et al} \cite{sam} \\ 
GRASP2b: present calculations from the {\sc grasp} code with 470 levels\\
GRASP3: present calculations from the {\sc grasp} code with 3990 levels\\
Ratio: ratio of velocity and length forms of A- values \\																       
}															       
\end{flushleft} 

\newpage
\clearpage

\begin{flushleft}
Table 11. Comparison of radiative rates (A- values, s$^{-1}$) for E1 transitions among levels of  the ground state 4s$^2$4p$^5$ $^2$P$^o_{3/2,1/2}$ and 4s4p$^6$ and 4s$^2$4p$^4$4d configurations of Mo VIII -- see Table 6 for level definitions.  $a{\pm}b \equiv a{\times}$10$^{{\pm}b}$.
\end{flushleft}
{\small
\begin{tabular}{rrllllllll} \hline
 & & & & & & & &  \\
 \multicolumn{2}{c}{Transition} & GRASP2a  & GRASP2b & \multicolumn{2}{c}{GRASP3} \\ \hline
   I &     J &     A   &       A     &     A     &   f       & Ratio \\
\hline  
     1  &  3 & 1.29+09  & 2.5432+09  & 2.1245+09 & 2.5943-02 & 4.6-1 \\
     1  &  4 & 2.28+07  & 2.9581+07  & 3.4543+07 & 7.7738-04 & 9.2-1 \\
     1  &  6 & 2.13+07  & 2.3934+07  & 2.9526+07 & 4.3870-04 & 1.0-0 \\
     1  &  7 & 2.00+07  & 2.0959+07  & 2.5720+07 & 1.8739-04 & 9.8-1 \\
     1  & 10 & 2.51+07  & 2.6107+07  & 2.9036+07 & 1.8539-04 & 1.3-0 \\
     1  & 11 & 3.63+08  & 4.5188+07  & 8.3700+07 & 1.0580-03 & 9.5-1 \\
     1  & 12 & 6.58+07  & 3.6663+08  & 5.0141+08 & 9.4488-03 & 9.5-1 \\
     1  & 13 & 1.19+09  & 1.1856+09  & 1.3282+09 & 8.1761-03 & 7.8-1 \\
     1  & 14 & 1.22+09  & 1.0656+09  & 1.3902+09 & 1.7040-02 & 9.7-1 \\
     1  & 15 & 4.32+08  & 3.2722+08  & 3.6056+08 & 4.3446-03 & 8.8-1 \\
     1  & 17 & 3.09+08  & 2.2203+08  & 2.4601+08 & 4.3398-03 & 8.0-1 \\
     1  & 18 & 8.29+07  & 7.5243+07  & 4.6619+07 & 5.3723-04 & 1.3-0 \\
     1  & 19 & 4.74+08  & 3.2086+08  & 3.8150+08 & 6.4450-03 & 9.9-1 \\
     1  & 22 & 3.42+08  & 2.8357+08  & 5.6353+08 & 9.2641-03 & 9.9-1 \\
     1  & 23 & 6.00+08  & 5.8754+08  & 6.0305+08 & 9.0695-03 & 8.1-1 \\
     1  & 25 & 3.93+09  & 3.3730+09  & 7.2030+09 & 6.5992-02 & 9.8-1 \\
     1  & 26 & 3.18+08  & 4.5232+08  & 4.0447+09 & 5.3300-02 & 1.0-0 \\
     1  & 27 & 2.56+11  & 2.3491+11  & 1.9071+11 & 7.3791-01 & 8.8-1 \\   
     1  & 28 & 2.39+11  & 2.0401+11  & 1.8003+11 & 1.4182-00 & 9.3-1 \\
     1  & 29 & 2.91+11  & 3.3603+09  & 3.4261+10 & 1.2685-01 & 7.4-1 \\   
     1  & 30 & 4.80+09  & 2.6848+11  & 2.3757+11 & 2.6875-00 & 9.0-1 \\
     1  & 31 & 8.74+09  & 9.1757+09  & 9.3027+09 & 6.4468-02 & 8.8-1 \\
     2  &  3 & 5.44+08  & 1.0480+09  & 9.0658+08 & 2.6857-02 & 4.2-1 \\
     2  &  6 & 1.75+06  & 2.7579+05  & 6.0372+05 & 2.0822-05 & 1.2-0 \\
     2  &  7 & 1.40+07  & 1.3921+07  & 1.7026+07 & 2.8753-04 & 1.2-0 \\
     2  & 10 & 1.06+08  & 1.1101+08  & 1.1742+08 & 1.7208-03 & 1.2-0 \\
     2  & 11 & 4.76+07  & 7.2300+07  & 7.4734+07 & 2.1668-03 & 1.1-0 \\
     2  & 13 & 7.85+07  & 1.0526+08  & 8.3988+07 & 1.1837-03 & 6.4-1 \\
     2  & 14 & 1.99+06  & 3.0099+06  & 7.7973+04 & 2.1873-06 & 1.0+1 \\
     2  & 15 & 5.42+08  & 4.1898+08  & 5.6237+08 & 1.5490-02 & 1.0-0 \\
     2  & 18 & 1.43+07  & 4.9598+06  & 9.8751+06 & 2.5935-04 & 6.3-1 \\
     2  & 25 & 3.26+09  & 2.5724+09  & 1.4586+09 & 3.0014-02 & 7.5-1 \\
     2  & 27 & 5.79+09  & 7.7879+09  & 1.7297+10 & 1.4887-01 & 1.1-0 \\
     2  & 28 & 3.16+09  & 2.2314+09  & 2.7063+09 & 4.7470-02 & 1.0-0 \\
     2  & 29 & 2.32+11  & 2.0141+11  & 1.7509+11 & 1.4386-00 & 8.7-1 \\   
     2  & 31 & 2.76+11  & 2.5240+11  & 2.2654+11 & 3.4718-00 & 9.1-1 \\ 
\hline            								                	 
\end{tabular}   								   					       
			      							   					       
\begin{flushleft}													       
{\small
GRASP2a: Singh {\em et al} \cite{sam} \\ 
GRASP2b: present calculations from the {\sc grasp} code with 470 levels\\
GRASP3: present calculations from the {\sc grasp} code with 3990 levels\\
Ratio: ratio of velocity and length forms of A- values \\																       
}															       
\end{flushleft} 
 
 \newpage
\clearpage

\begin{flushleft}
Table 12. Comparison of lifetimes ($\tau$, s)  for the lowest 31 levels of Sr IV.  $a{\pm}b \equiv a{\times}$10$^{{\pm}b}$.
\end{flushleft}
{\small
\begin{tabular}{rlllllllll} \hline
 & & & & & & & &  \\
Index  & Configuration         & Level              & GRASP2a & GRASP2b & GRASP3  & GRASP2b (dominant A- values, s$^{-1}$)    \\
& & & & & & & &   \\ \hline  
& & & & & & & &   \\
  1  & 4s$^2$4p$^5$	       &  $^2$P$^o_{3/2}$   & ....... & ........ & ........  &         ........                                        \\ 
  2  & 4s$^2$4p$^5$	       &  $^2$P$^o_{1/2}$   & 6.18-02 & 7.112-02 & 6.836-02  &     1  -  2     M1 = 1.405+01			       \\
  3  & 4s4p$^6$ 	       &  $^2$S$  _{1/2}$   & 1.71-09 & 1.870-09 & 1.884-09  &     1  -  3     E1 = 3.634+08, 2  -  3, E1 = 1.714+08   \\
  4  & 4s$^2$4p$^4$($^3$P)4d   &  $^4$D$  _{7/2}$   & 1.90-01 & 1.484-01 & 1.193-01  &     1  -  4     M2 = 6.740+00			       \\
  5  & 4s$^2$4p$^4$($^3$P)4d   &  $^4$D$  _{5/2}$   & 1.17-07 & 1.691-07 & 9.774-08  &     1  -  5     E1 = 5.914+06			       \\
  6  & 4s$^2$4p$^4$($^3$P)4d   &  $^4$D$  _{3/2}$   & 3.64-07 & 4.123-07 & 3.535-07  &     1  -  6     E1 = 2.402+06			       \\
  7  & 4s$^2$4p$^4$($^3$P)4d   &  $^4$D$  _{1/2}$   & 1.39-06 & 5.620-07 & 1.308-06  &     1  -  7     E1 = 8.062+05, 2  -  7, E1 = 9.732+05   \\
  8  & 4s$^2$4p$^4$($^3$P)4d   &  $^4$F$  _{9/2}$   & 7.45-01 & 7.697-01 & 7.896-01  &     4  -  8     M1 = 1.287+00			       \\
  9  & 4s$^2$4p$^4$($^3$P)4d   &  $^4$F$  _{7/2}$   & 1.73-00 & 2.984-01 & 2.957-01  &     4  -  9     M1 = 1.231+00, 8  -  9, M1 = 1.374+00   \\
 10  & 4s$^2$4p$^4$($^1$D)4d   &  $^2$P$  _{1/2}$   & 1.03-06 & 3.376-08 & 2.445-07  &     2  - 10     E1 = 2.516+07			       \\
 11  & 4s$^2$4p$^4$($^3$P)4d   &  $^4$F$  _{5/2}$   & 1.59-08 & 2.133-08 & 1.232-08  &     1  - 11     E1 = 4.687+07			       \\
 12  & 4s$^2$4p$^4$($^3$P)4d   &  $^4$F$  _{3/2}$   & 2.01-07 & 3.397-08 & 2.516-08  &     1  - 12     E1 = 6.598+06, 2  - 12, E1 = 2.284+07   \\
 13  & 4s$^2$4p$^4$($^3$P)4d   &  $^4$P$  _{1/2}$   & 4.60-09 & 6.673-09 & 3.819-09  &     1  - 13     E1 = 1.434+08			       \\
 14  & 4s$^2$4p$^4$($^3$P)4d   &  $^4$P$  _{3/2}$   & 5.75-09 & 5.659-09 & 5.661-09  &     1  - 14     E1 = 1.622+08			       \\
 15  & 4s$^2$4p$^4$($^1$D)4d   &  $^2$D$  _{3/2}$   & 5.28-09 & 3.047-09 & 2.111-09  &     1  - 15     E1 = 1.439+08, 2  - 15, E1 = 1.843+08   \\
 16  & 4s$^2$4p$^4$($^3$P)4d   &  $^2$F$  _{7/2}$   & 8.43-01 & 7.657-02 & 7.731-02  &     4  - 16     M1 = 5.728+00, 8  - 16, M1 = 3.032+00   \\
 17  & 4s$^2$4p$^4$($^3$P)4d   &  $^4$P$  _{5/2}$   & 1.71-08 & 2.011-08 & 1.315-08  &     1  - 17     E1 = 4.973+07			       \\
 18  & 4s$^2$4p$^4$($^1$D)4d   &  $^2$P$  _{3/2}$   & 1.59-08 & 8.926-09 & 3.277-08  &     1  - 18     E1 = 7.491+07, 2  - 18, E1 = 3.712+07   \\
 19  & 4s$^2$4p$^4$($^1$D)4d   &  $^2$D$  _{5/2}$   & 3.88-09 & 4.372-09 & 4.290-09  &     1  - 19     E1 = 2.287+08			       \\
 20  & 4s$^2$4p$^4$($^3$P)4d   &  $^2$F$  _{5/2}$   & 1.05-08 & 8.131-09 & 6.120-09  &     1  - 20     E1 = 1.230+08			       \\
 21  & 4s$^2$4p$^4$($^1$D)4d   &  $^2$G$  _{9/2}$   & 7.11-01 & 5.270-02 & 5.642-02  &     8  - 21     M1 = 1.501+01			       \\
 22  & 4s$^2$4p$^4$($^1$D)4d   &  $^2$G$  _{7/2}$   & 1.66-00 & 6.495-02 & 6.780-02  &     9  - 22     M1 = 6.240+00			       \\
 23  & 4s$^2$4p$^4$($^1$D)4d   &  $^2$F$  _{5/2}$   & 5.32-09 & 7.177-09 & 4.276-09  &     1  - 23     E1 = 1.393+08			       \\
 24  & 4s$^2$4p$^4$($^1$D)4d   &  $^2$F$  _{7/2}$   & 3.83-00 & 2.309-02 & 2.357-02  &     4  - 24     M1 = 1.547+01, 8  - 24, M1 = 1.351+01   \\
 25  & 4s$^2$4p$^4$($^1$S)4d   &  $^2$D$  _{3/2}$   & 1.93-09 & 5.031-10 & 4.380-10  &     1  - 25     E1 = 4.710+08, 2  - 25, E1 = 1.517+09   \\
 26  & 4s$^2$4p$^4$($^1$S)4d   &  $^2$D$  _{5/2}$   & 3.04-09 & 3.997-09 & 6.305-10  &     1  - 26     E1 = 2.502+08			       \\
 27  & 4s$^2$4p$^4$($^3$P)4d   &  $^2$P$  _{3/2}$   & 1.17-11 & 1.403-11 & 1.713-11  &     1  - 27     E1 = 6.951+10			       \\
 28  & 4s$^2$4p$^4$($^3$P)4d   &  $^2$P$  _{1/2}$   & 1.82-11 & 1.383-11 & 1.714-11  &     1  - 28     E1 = 3.877+10, 2  - 28, E1 = 3.352+10   \\
 29  & 4s$^2$4p$^4$($^3$P)4d   &  $^2$D$  _{5/2}$   & 8.37-12 & 1.087-11 & 1.725-11  &     1  - 29     E1 = 9.197+10			       \\
 30  & 4s$^2$4p$^4$($^1$D)4d   &  $^2$S$  _{1/2}$   & 1.49-11 & 1.001-11 & 1.383-11  &     1  - 30     E1 = 5.427+10, 2  - 30, E1 = 4.567+10   \\
 31  & 4s$^2$4p$^4$($^3$P)4d   &  $^2$D$  _{3/2}$   & 8.52-12 & 1.109-11 & 1.377-11  &     2  - 31     E1 = 8.557+10			       \\
& & & & & & & & \\ \hline            								                	 
\end{tabular}   								   					       
			      							   					       
\begin{flushleft}													       
{\small
GRASP2a: Singh{\em  et al} \cite{sam} \\ 
GRASP2b: present calculations from the {\sc grasp} code with 470 levels\\
GRASP3: present calculations from the {\sc grasp} code with 3990 levels\\
															       
}															       
\end{flushleft} 
 
 \newpage
\clearpage

\begin{flushleft}
Table 13. Comparison of lifetimes ($\tau$, s)  for the lowest 31 levels of Y V.  $a{\pm}b \equiv a{\times}$10$^{{\pm}b}$.
\end{flushleft}
{\small
\begin{tabular}{rllllllll} \hline
 & & & & & & & &  \\
Index  & Configuration         & Level              & GRASP2a & GRASP2b & GRASP3  & GRASP2b (dominant A- values, s$^{-1}$)    \\
& & & & & & & &   \\ \hline  
& & & & & & & &   \\
  1  & 4s$^2$4p$^5$	       &  $^2$P$^o_{3/2}$  & ....... & ........ & ........  &	      ........  				                             \\ 
  2  & 4s$^2$4p$^5$	       &  $^2$P$^o_{1/2}$  & 2.90-02 & 3.278-02 & 3.148-02  &  1 -   2     M1 = 3.047+01						     \\ 
  3  & 4s4p$^6$ 	       &  $^2$S$  _{1/2}$  & 9.67-10 & 9.305-10 & 1.052-09  &  1 -   3     E1 = 7.418+08, 2  -  3  E1 = 3.329+08			     \\ 
  4  & 4s$^2$4p$^4$($^3$P)4d   &  $^4$D$  _{5/2}$  & 7.37-08 & 9.951-08 & 6.805-08  &  1 -   4     E1 = 1.005+07						     \\ 
  5  & 4s$^2$4p$^4$($^3$P)4d   &  $^4$D$  _{7/2}$  & 9.74-02 & 7.091-02 & 6.619-02  &  1 -   5     M2 = 1.410+01						     \\ 
  6  & 4s$^2$4p$^4$($^3$P)4d   &  $^4$D$  _{3/2}$  & 1.78-07 & 1.958-07 & 1.726-07  &  1 -   6     E1 = 5.098+06						     \\ 
  7  & 4s$^2$4p$^4$($^3$P)4d   &  $^4$D$  _{1/2}$  & 4.53-07 & 2.087-07 & 2.694-07  &  1 -   7     E1 = 2.309+06, 2  -  7  E1 = 2.482+06			     \\ 
  8  & 4s$^2$4p$^4$($^3$P)4d   &  $^4$F$  _{9/2}$  & 1.04+02 & 4.485-01 & 4.804-01  &  5 -   8     M1 = 2.219+00					             \\ 
  9  & 4s$^2$4p$^4$($^3$P)4d   &  $^4$F$  _{7/2}$  & 7.22-01 & 1.555-01 & 1.573-01  &  1 -   9     M2 = 1.731+00, 5  -  9  M1 = 2.122+00,  8 - 9  M1 = 2.457+00      \\ 
 10  & 4s$^2$4p$^4$($^1$D)4d   &  $^2$P$  _{1/2}$  & 3.28-07 & 1.727-08 & 3.383-08  &  2 -  10     E1 = 4.874+07						     \\ 
 11  & 4s$^2$4p$^4$($^3$P)4d   &  $^4$F$  _{5/2}$  & 9.53-09 & 1.197-08 & 7.447-09  &  1 -  11     E1 = 8.356+07						     \\ 
 12  & 4s$^2$4p$^4$($^3$P)4d   &  $^4$F$  _{3/2}$  & 5.55-06 & 2.548-08 & 2.329-08  &  2 -  12     E1 = 3.396+07						     \\ 
 13  & 4s$^2$4p$^4$($^3$P)4d   &  $^4$P$  _{1/2}$  & 2.66-09 & 3.463-09 & 2.470-09  &  1 -  13     E1 = 2.711+08						     \\ 
 14  & 4s$^2$4p$^4$($^3$P)4d   &  $^4$P$  _{3/2}$  & 3.36-09 & 3.178-09 & 2.380-09  &  1 -  14     E1 = 2.929+08						     \\ 
 15  & 4s$^2$4p$^4$($^1$D)4d   &  $^2$D$  _{3/2}$  & 3.98-09 & 2.310-09 & 1.764-09  &  1 -  15     E1 = 2.045+08, 2  - 15  E1 = 2.284+08			     \\ 
 16  & 4s$^2$4p$^4$($^3$P)4d   &  $^2$F$  _{7/2}$  & 5.50-01 & 3.952-02 & 4.055-02  &  5 -  16     M1 = 1.057+01, 8  - 16  M1 = 6.762+00			     \\ 
 17  & 4s$^2$4p$^4$($^3$P)4d   &  $^4$P$  _{5/2}$  & 1.17-08 & 1.278-08 & 8.082-09  &  1 -  17     E1 = 7.826+07						     \\ 
 18  & 4s$^2$4p$^4$($^1$D)4d   &  $^2$P$  _{3/2}$  & 1.90-08 & 1.237-08 & 4.151-08  &  1 -  18     E1 = 6.462+07, 2  - 18  E1 = 1.622+07			     \\ 
 19  & 4s$^2$4p$^4$($^1$D)4d   &  $^2$D$  _{5/2}$  & 3.84-09 & 4.094-09 & 4.029-09  &  1 -  19     E1 = 2.443+08						     \\ 
 20  & 4s$^2$4p$^4$($^1$D)4d   &  $^2$G$  _{9/2}$  & 4.77+01 & 2.856-02 & 2.994-02  &  8 -  20     M1 = 2.795+01						     \\ 
 21  & 4s$^2$4p$^4$($^1$D)4d   &  $^2$G$  _{7/2}$  & 7.23-01 & 3.336-02 & 3.447-02  &  5 -  21     M1 = 6.539+00, 9  - 21  M1 = 1.134+01			     \\ 
 22  & 4s$^2$4p$^4$($^3$P)4d   &  $^2$F$  _{5/2}$  & 1.11-08 & 8.379-09 & 4.354-09  &  1 -  22     E1 = 1.193+08						     \\ 
 23  & 4s$^2$4p$^4$($^1$D)4d   &  $^2$F$  _{5/2}$  & 3.62-09 & 4.654-09 & 2.985-09  &  1 -  23     E1 = 2.149+08						     \\ 
 24  & 4s$^2$4p$^4$($^1$D)4d   &  $^2$F$  _{7/2}$  & 2.93-00 & 1.258-02 & 1.300-02  &  5 -  24     M1 = 2.711+01, 8  - 24  M1 = 2.758+01			     \\ 
 25  & 4s$^2$4p$^4$($^1$S)4d   &  $^2$D$  _{3/2}$  & 1.11-09 & 4.283-10 & 2.708-10  &  1 -  25     E1 = 8.617+08, 2  - 25  E1 = 1.473+09			     \\ 
 26  & 4s$^2$4p$^4$($^1$S)4d   &  $^2$D$  _{5/2}$  & 1.70-08 & 5.369-08 & 6.716-10  &  1 -  26     E1 = 1.863+07						     \\ 
 27  & 4s$^2$4p$^4$($^3$P)4d   &  $^2$P$  _{3/2}$  & 8.35-12 & 9.650-12 & 9.174-12  &  1 -  27     E1 = 1.014+11						     \\ 
 28  & 4s$^2$4p$^4$($^3$P)4d   &  $^2$P$  _{1/2}$  & 8.37-12 & 9.143-12 & 9.153-12  &  1 -  28     E1 = 8.286+10, 2  - 28  E1 = 2.651+10			     \\ 
 29  & 4s$^2$4p$^4$($^3$P)4d   &  $^2$D$  _{5/2}$  & 5.93-12 & 7.416-12 & 8.183-12  &  1 -  29     E1 = 1.349+11						     \\ 
 30  & 4s$^2$4p$^4$($^1$D)4d   &  $^2$S$  _{1/2}$  & 2.32-11 & 7.725-12 & 1.016-11  &  1 -  30     E1 = 4.503+10, 2  - 30  E1 = 8.443+10			     \\ 
 31  & 4s$^2$4p$^4$($^3$P)4d   &  $^2$D$  _{3/2}$  & 6.05-12 & 7.572-12 & 8.006-12  &  2 -  31     E1 = 1.258+11						     \\ 
& & & & & & & & \\ \hline            								                	 
\end{tabular}   								   					       
			      							   					       
\begin{flushleft}													       
{\small
GRASP2a: Singh {\em et al} \cite{sam} \\ 
GRASP2b: present calculations from the {\sc grasp} code with 470 levels\\
GRASP3: present calculations from the {\sc grasp} code with 3990 levels\\
															       
}															       
\end{flushleft} 

\newpage
\clearpage

\begin{flushleft}
Table 14. Comparison of lifetimes ($\tau$, s)  for the lowest 31 levels of Zr VI.  $a{\pm}b \equiv a{\times}$10$^{{\pm}b}$.
\end{flushleft}
{\small
\begin{tabular}{rllllllll} \hline
 & & & & & & & &  \\
Index  & Configuration         & Level              & GRASP2a & GRASP2b & GRASP3  & GRASP2b (dominant A- values, s$^{-1}$)    \\
& & & & & & & &   \\ \hline  
& & & & & & & &   \\
  1  & 4s$^2$4p$^5$	       &  $^2$P$^o_{3/2}$   & ....... & ........ & ........  &         ........ 								\\ 
  2  & 4s$^2$4p$^5$	       &  $^2$P$^o_{1/2}$   & 1.47-02 & 1.630-02 & 1.583-02  &  1  -  2  M1 =  6.126+01 							\\
  3  & 4s4p$^6$ 	       &  $^2$S$  _{1/2}$   & 6.31-10 & 5.650-10 & 6.560-10  &  1  -  3  E1 =  1.234+09,  2  -  3  E1 =  5.359+08				\\
  4  & 4s$^2$4p$^4$($^3$P)4d   &  $^4$D$  _{5/2}$   & 5.03-08 & 6.497-08 & 4.924-08  &  1  -  4  E1 =  1.539+07 							\\
  5  & 4s$^2$4p$^4$($^3$P)4d   &  $^4$D$  _{7/2}$   & 3.59-02 & 3.966-02 & 3.951-02  &  1  -  5  M2 =  2.521+01 							\\
  6  & 4s$^2$4p$^4$($^3$P)4d   &  $^4$D$  _{3/2}$   & 1.00-07 & 1.076-07 & 9.077-08  &  1  -  6  E1 =  9.293+06 							\\
  7  & 4s$^2$4p$^4$($^3$P)4d   &  $^4$D$  _{1/2}$   & 1.91-07 & 9.679-08 & 9.501-08  &  1  -  7  E1 =  5.346+06,  2  -  7  E1 =  4.986+06				\\
  8  & 4s$^2$4p$^4$($^3$P)4d   &  $^4$F$  _{9/2}$   & 9.02+01 & 2.876-01 & 3.117-01  &  5  -  8  M1 =  3.468+00 							\\
  9  & 4s$^2$4p$^4$($^3$P)4d   &  $^4$F$  _{7/2}$   & 1.80-01 & 8.920-02 & 9.182-02  &  1  -  9  M2 =  3.848+00,  5  -  9  M1 =  3.250+00,  8  -  9  M1 =  3.943+00	\\
 10  & 4s$^2$4p$^4$($^1$D)4d   &  $^2$P$  _{1/2}$   & 1.81-07 & 1.150-08 & 4.717-09  &  2  - 10  E1 =  7.218+07 							\\
 11  & 4s$^2$4p$^4$($^3$P)4d   &  $^4$F$  _{5/2}$   & 5.83-09 & 7.043-09 & 1.401-08  &  1  - 11  E1 =  1.420+08 							\\
 12  & 4s$^2$4p$^4$($^3$P)4d   &  $^4$F$  _{3/2}$   & 6.16-08 & 2.081-08 & 1.916-08  &  2  - 12  E1 =  4.704+07 							\\
 13  & 4s$^2$4p$^4$($^3$P)4d   &  $^4$P$  _{1/2}$   & 1.62-09 & 1.974-09 & 1.601-09  &  1  - 13  E1 =  4.700+08 							\\
 14  & 4s$^2$4p$^4$($^3$P)4d   &  $^4$P$  _{3/2}$   & 2.21-09 & 1.966-09 & 1.421-09  &  1  - 14  E1 =  4.884+08 							\\
 15  & 4s$^2$4p$^4$($^1$D)4d   &  $^2$D$  _{3/2}$   & 3.16-09 & 1.882-09 & 1.529-09  &  1  - 15  E1 =  2.519+08,  2  - 15  E1 =  2.795+08				\\
 16  & 4s$^2$4p$^4$($^3$P)4d   &  $^2$F$  _{7/2}$   & 4.11-01 & 2.171-02 & 2.241-02  &  5  - 16  M1 =  1.845+01,  8  - 16  M1 =  1.447+01				\\
 17  & 4s$^2$4p$^4$($^3$P)4d   &  $^4$P$  _{5/2}$   & 8.09-09 & 8.645-09 & 6.409-09  &  1  - 17  E1 =  1.157+08 							\\
 18  & 4s$^2$4p$^4$($^1$D)4d   &  $^2$P$  _{3/2}$   & 1.96-08 & 1.402-08 & 3.474-08  &  1  - 18  E1 =  6.290+07 							\\
 19  & 4s$^2$4p$^4$($^1$D)4d   &  $^2$D$  _{5/2}$   & 3.69-09 & 3.912-09 & 3.794-09  &  1  - 19  E1 =  2.556+08 							\\
 20  & 4s$^2$4p$^4$($^1$D)4d   &  $^2$G$  _{7/2}$   & 2.42-01 & 1.810-02 & 1.879-02  &  5  - 20  M1 =  1.393+01,  9  - 20  M1 =  2.029+01				\\
 21  & 4s$^2$4p$^4$($^1$D)4d   &  $^2$G$  _{9/2}$   & 9.88+00 & 1.609-02 & 1.686-02  &  8  - 21  M1 =  4.981+01 							\\
 22  & 4s$^2$4p$^4$($^3$P)4d   &  $^2$F$  _{5/2}$   & 8.90-09 & 6.831-09 & 3.300-09  &  1  - 22  E1 =  1.464+08 							\\
 23  & 4s$^2$4p$^4$($^1$D)4d   &  $^2$F$  _{5/2}$   & 2.55-09 & 3.187-09 & 2.525-09  &  1  - 23  E1 =  3.138+08 							\\
 24  & 4s$^2$4p$^4$($^1$D)4d   &  $^2$F$  _{7/2}$   & 2.26-00 & 7.230-03 & 7.502-03  &  5  - 24  M1 =  4.515+01,  8  - 24  M1 =  5.213+01				\\
 25  & 4s$^2$4p$^4$($^1$S)4d   &  $^2$D$  _{3/2}$   & 6.89-10 & 3.240-10 & 2.128-10  &  1  - 25  E1 =  1.435+09,  2  - 25  E1 =  1.651+09				\\
 26  & 4s$^2$4p$^4$($^1$S)4d   &  $^2$D$  _{5/2}$   & 1.26-04 & 4.573-08 & 4.738-10  &  1  - 26  E1 =  2.187+07 							\\
 27  & 4s$^2$4p$^4$($^3$P)4d   &  $^2$P$  _{3/2}$   & 6.45-12 & 7.285-12 & 7.821-12  &  1  - 27  E1 =  1.348+11 							\\
 28  & 4s$^2$4p$^4$($^1$D)4d   &  $^2$S$  _{1/2}$   & 5.08-12 & 6.380-12 & 7.341-12  &  1  - 28  E1 =  1.533+11 							\\
 29  & 4s$^2$4p$^4$($^3$P)4d   &  $^2$P$  _{1/2}$   & 1.65-10 & 6.637-12 & 6.219-12  &  2  - 29  E1 =  1.401+11 							\\
 30  & 4s$^2$4p$^4$($^3$P)4d   &  $^2$D$  _{5/2}$   & 4.60-12 & 5.588-12 & 6.190-12  &  1  - 30  E1 =  1.790+11 							\\
 31  & 4s$^2$4p$^4$($^3$P)4d   &  $^2$D$  _{3/2}$   & 4.70-12 & 5.714-12 & 6.221-12  &  2  - 31  E1 =  1.674+11 							\\
& & & & & & & & \\ \hline            										
\end{tabular}   								   					       
			      							   					       
\begin{flushleft}													       
{\small
GRASP2a: Singh {\em et al} \cite{sam} \\ 
GRASP2b: present calculations from the {\sc grasp} code with 470 levels\\
GRASP3: present calculations from the {\sc grasp} code with 3990 levels\\
															       
}															       
\end{flushleft} 

\newpage
\clearpage

\begin{flushleft}
Table 15. Comparison of lifetimes ($\tau$, s)  for the lowest 31 levels of Nb VII.  $a{\pm}b \equiv a{\times}$10$^{{\pm}b}$.
\end{flushleft}
{\small
\begin{tabular}{rllllllll} \hline
 & & & & & & & &  \\
Index  & Configuration         & Level              & GRASP2a & GRASP2b & GRASP3  & GRASP2b (dominant A- values, s$^{-1}$)    \\
& & & & & & & &   \\ \hline  
& & & & & & & &   \\
  1  & 4s$^2$4p$^5$	       &  $^2$P$^o_{3/2}$   &  ....... & ........ & ........  & 	........							      \\ 
  2  & 4s$^2$4p$^5$	       &  $^2$P$^o_{1/2}$   &  7.84-03 & 8.602-03 & 8.426-03  &  1  -  2  M1 = 1.161+02 						      \\
  3  & 4s4p$^6$ 	       &  $^2$S$  _{1/2}$   &  4.48-10 & 3.829-10 & 4.508-10  &  1  -  3  E1 = 1.836+09, 2  -  3 E1 =	 7.757+08			      \\
  4  & 4s$^2$4p$^4$($^3$P)4d   &  $^4$D$  _{5/2}$   &  3.62-08 & 4.560-08 & 3.711-08  &  1  -  4  E1 = 2.193+07 						      \\
  5  & 4s$^2$4p$^4$($^3$P)4d   &  $^4$D$  _{7/2}$   &  3.38+02 & 2.458-02 & 2.555-02  &  1  -  5  M2 = 4.068+01 						      \\
  6  & 4s$^2$4p$^4$($^3$P)4d   &  $^4$D$  _{3/2}$   &  6.18-08 & 6.459-08 & 5.290-08  &  1  -  6  E1 = 1.542+07 						      \\
  7  & 4s$^2$4p$^4$($^3$P)4d   &  $^4$D$  _{1/2}$   &  9.19-08 & 5.060-08 & 4.391-08  &  1  -  7  E1 = 1.102+07, 2  -  7 E1 =	 8.749+06			      \\
  8  & 4s$^2$4p$^4$($^3$P)4d   &  $^4$F$  _{9/2}$   &  1.84-01 & 1.977-01 & 2.165-01  &  5  -  8  M1 = 5.050+00 						      \\
  9  & 4s$^2$4p$^4$($^3$P)4d   &  $^4$F$  _{7/2}$   &  3.86+00 & 5.568-02 & 5.843-02  &  1  -  9  M2 = 7.535+00, 5  -  9 M1 =	 4.524+00, 8  -  9  M1 =  5.677+00    \\
 10  & 4s$^2$4p$^4$($^1$D)4d   &  $^2$P$  _{1/2}$   &  1.30-07 & 8.766-09 & 8.913-09  &  2  - 10  E1 = 9.337+07 						      \\
 11  & 4s$^2$4p$^4$($^3$P)4d   &  $^4$F$  _{5/2}$   &  3.66-09 & 4.311-09 & 3.037-09  &  1  - 11  E1 = 2.320+08 						      \\
 12  & 4s$^2$4p$^4$($^3$P)4d   &  $^4$F$  _{3/2}$   &  8.98-09 & 1.552-08 & 1.204-08  &  2  - 12  E1 = 6.066+07 						      \\
 13  & 4s$^2$4p$^4$($^3$P)4d   &  $^4$P$  _{1/2}$   &  1.03-09 & 1.205-09 & 1.052-09  &  1  - 13  E1 = 7.649+08 						      \\
 14  & 4s$^2$4p$^4$($^3$P)4d   &  $^4$P$  _{3/2}$   &  1.70-09 & 1.306-09 & 9.735-10  &  1  - 14  E1 = 7.539+08 						      \\
 15  & 4s$^2$4p$^4$($^1$D)4d   &  $^2$D$  _{3/2}$   &  2.55-09 & 1.581-09 & 1.296-09  &  1  - 15  E1 = 2.880+08, 2 -  15 E1 =	 3.445+08			      \\
 16  & 4s$^2$4p$^4$($^3$P)4d   &  $^2$F$  _{7/2}$   &  2.39-01 & 1.230-02 & 1.267-02  &  5  - 16  M1 = 3.116+01, 8 -  16 M1 =	 2.958+01			      \\
 17  & 4s$^2$4p$^4$($^3$P)4d   &  $^4$P$  _{5/2}$   &  5.74-09 & 6.149-09 & 5.163-09  &  1  - 17  E1 = 1.626+08 						      \\
 18  & 4s$^2$4p$^4$($^1$D)4d   &  $^2$P$  _{3/2}$   &  1.83-08 & 1.378-08 & 2.510-08  &  1  - 18  E1 = 6.703+07 						      \\
 19  & 4s$^2$4p$^4$($^1$D)4d   &  $^2$D$  _{5/2}$   &  3.37-09 & 3.601-09 & 3.288-09  &  1  - 19  E1 = 2.777+08 						      \\
 20  & 4s$^2$4p$^4$($^1$D)4d   &  $^2$G$  _{7/2}$   &  7.27-02 & 1.032-02 & 1.076-02  &  5  - 20  M1 = 2.681+01, 9 -  20 M1 =	 3.567+01			      \\
 21  & 4s$^2$4p$^4$($^1$D)4d   &  $^2$G$  _{9/2}$   &  8.49-02 & 9.372-03 & 9.802-03  &  8  - 21  M1 = 8.564+01 						      \\
 22  & 4s$^2$4p$^4$($^3$P)4d   &  $^2$F$  _{5/2}$   &  6.36-09 & 5.001-09 & 2.423-09  &  1  - 22  E1 = 1.999+08 						      \\
 23  & 4s$^2$4p$^4$($^1$D)4d   &  $^2$F$  _{5/2}$   &  1.85-09 & 2.284-09 & 2.053-09  &  1  - 23  E1 = 4.379+08 						      \\
 24  & 4s$^2$4p$^4$($^1$D)4d   &  $^2$F$  _{7/2}$   &  5.67-02 & 4.334-03 & 4.507-03  &  5  - 24  M1 = 7.219+01, 8 -  24 M1 =	 9.290+01			      \\
 25  & 4s$^2$4p$^4$($^1$S)4d   &  $^2$D$  _{3/2}$   &  4.46-10 & 2.347-10 & 1.580-10  &  1  - 25  E1 = 2.247+09, 2 -  25 E1 =	 2.013+09			      \\
 26  & 4s$^2$4p$^4$($^1$S)4d   &  $^2$D$  _{5/2}$   &  2.20-08 & 5.985-09 & 3.420-10  &  1  - 26  E1 = 1.671+08 						      \\
 27  & 4s$^2$4p$^4$($^1$D)4d   &  $^2$S$  _{1/2}$   &  4.10-12 & 4.945-12 & 5.865-12  &  1  - 27  E1 = 2.008+11 						      \\
 28  & 4s$^2$4p$^4$($^3$P)4d   &  $^2$P$  _{3/2}$   &  5.24-12 & 5.828-12 & 6.450-12  &  1  - 28  E1 = 1.691+11 						      \\
 29  & 4s$^2$4p$^4$($^3$P)4d   &  $^2$P$  _{1/2}$   &  5.84-09 & 4.472-12 & 5.359-12  &  1  - 29  E1 = 2.236+11 						      \\
 30  & 4s$^2$4p$^4$($^3$P)4d   &  $^2$D$  _{5/2}$   &  3.75-12 & 5.717-12 & 5.005-12  &  2  - 30  E1 = 1.749+11 						      \\
 31  & 4s$^2$4p$^4$($^3$P)4d   &  $^2$D$  _{3/2}$   &  3.84-12 & 4.581-12 & 5.039-12  &  2  - 31  E1 = 2.097+11 						      \\
& & & & & & & & \\ \hline            										
\end{tabular}   								   					       
			      							   					       
\begin{flushleft}													       
{\small
GRASP2a: Singh {\em et al} \cite{sam} \\ 
GRASP2b: present calculations from the {\sc grasp} code with 470 levels\\
GRASP3: present calculations from the {\sc grasp} code with 3990 levels\\
															       
}															       
\end{flushleft} 

\newpage
\clearpage

\begin{flushleft}
Table 16. Comparison of lifetimes ($\tau$, s)  for the lowest 31 levels of Mo VIII.  $a{\pm}b \equiv a{\times}$10$^{{\pm}b}$.
\end{flushleft}
{\small
\begin{tabular}{rllllllll} \hline
 & & & & & & & &  \\
Index  & Configuration         & Level              & GRASP2a & GRASP2b & GRASP3  & GRASP2b (dominant A- values, s$^{-1}$)    \\
& & & & & & & &   \\ \hline  
& & & & & & & &   \\
  1  & 4s$^2$4p$^5$	       &  $^2$P$^o_{3/2}$   &  ....... & ........ & ........  & 	........							\\ 
  2  & 4s$^2$4p$^5$	       &  $^2$P$^o_{1/2}$   &  4.38-03 & 4.762-03 & 4.692-03  &  1  -  2 M1 = 2.096+02  						\\
  3  & 4s4p$^6$ 	       &  $^2$S$  _{1/2}$   &  3.36-10 & 2.785-10 & 3.299-10  &  1  -  3 E1 = 2.543+09, 2  -  3 E1 = 1.048+09				\\
  4  & 4s$^2$4p$^4$($^3$P)4d   &  $^4$D$  _{5/2}$   &  2.72-08 & 3.381-08 & 2.895-08  &  1  -  4 E1 = 2.958+07  						\\
  5  & 4s$^2$4p$^4$($^3$P)4d   &  $^4$D$  _{7/2}$   &  6.75+01 & 1.638-02 & 1.752-02  &  1  -  5 M2 = 6.103+01  						\\
  6  & 4s$^2$4p$^4$($^3$P)4d   &  $^4$D$  _{3/2}$   &  4.06-08 & 4.131-08 & 3.319-08  &  1  -  6 E1 = 2.393+07  						\\
  7  & 4s$^2$4p$^4$($^3$P)4d   &  $^4$D$  _{1/2}$   &  4.83-08 & 2.867-08 & 2.339-08  &  1  -  7 E1 = 2.096+07, 2  -  7 E1 = 1.392+07				\\
  8  & 4s$^2$4p$^4$($^3$P)4d   &  $^4$F$  _{9/2}$   &  1.32-01 & 1.433-01 & 1.583-01  &  5  -  8 M1 = 6.968+00  						\\
  9  & 4s$^2$4p$^4$($^3$P)4d   &  $^4$F$  _{7/2}$   &  2.99+00 & 3.734-02 & 3.994-02  &  1  -  9 M2 = 1.330+01, 5  -  9 M1 = 5.843+00, 8  -  9 M1 = 7.362+00	\\
 10  & 4s$^2$4p$^4$($^1$D)4d   &  $^2$P$  _{1/2}$   &  1.15-07 & 7.293-09 & 6.828-09  &  2  - 10 E1 = 1.110+08  						\\
 12  & 4s$^2$4p$^4$($^3$P)4d   &  $^4$F$  _{3/2}$   &  3.61-09 & 8.512-09 & 6.312-09  &  1  - 11 E1 = 4.519+07, 2  - 11 E1 = 7.230+07				\\
 11  & 4s$^2$4p$^4$($^3$P)4d   &  $^4$F$  _{5/2}$   &  2.35-09 & 2.728-09 & 1.994-09  &  1  - 12 E1 = 3.666+08  						\\
 13  & 4s$^2$4p$^4$($^3$P)4d   &  $^4$P$  _{1/2}$   &  6.82-10 & 7.747-10 & 7.081-10  &  1  - 13 E1 = 1.186+09  						\\
 14  & 4s$^2$4p$^4$($^3$P)4d   &  $^4$P$  _{3/2}$   &  1.39-09 & 9.358-10 & 7.193-10  &  1  - 14 E1 = 1.066+09  						\\
 15  & 4s$^2$4p$^4$($^1$D)4d   &  $^2$D$  _{3/2}$   &  2.02-09 & 1.340-09 & 1.084-09  &  1  - 15 E1 = 3.272+08, 2  - 15 E1 = 4.190+08				\\
 16  & 4s$^2$4p$^4$($^3$P)4d   &  $^2$F$  _{7/2}$   &  1.58-01 & 7.107-03 & 7.278-03  &  5  - 16 M1 = 5.117+01, 8  - 16 M1 = 5.760+01				\\
 17  & 4s$^2$4p$^4$($^3$P)4d   &  $^4$P$  _{5/2}$   &  4.13-09 & 4.504-09 & 4.065-09  &  1  - 17 E1 = 2.220+08  						\\
 18  & 4s$^2$4p$^4$($^1$D)4d   &  $^2$P$  _{3/2}$   &  1.61-08 & 1.247-08 & 1.770-08  &  1  - 18 E1 = 7.524+07  						\\
 19  & 4s$^2$4p$^4$($^1$D)4d   &  $^2$D$  _{5/2}$   &  2.91-09 & 3.117-09 & 2.621-09  &  1  - 19 E1 = 3.209+08  						\\
 20  & 4s$^2$4p$^4$($^1$D)4d   &  $^2$G$  _{7/2}$   &  4.15-02 & 6.123-03 & 6.396-03  &  5  - 20 M1 = 4.821+01, 9  - 20 M1 = 6.120+01				\\
 21  & 4s$^2$4p$^4$($^1$D)4d   &  $^2$G$  _{9/2}$   &  4.68-02 & 5.617-03 & 5.859-03  &  8  - 21 M1 = 1.429+02  						\\
 22  & 4s$^2$4p$^4$($^3$P)4d   &  $^2$F$  _{5/2}$   &  4.37-09 & 3.526-09 & 1.775-09  &  1  - 22 E1 = 2.836+08  						\\
 23  & 4s$^2$4p$^4$($^1$D)4d   &  $^2$F$  _{5/2}$   &  1.39-09 & 1.702-09 & 1.658-09  &  1  - 23 E1 = 5.875+08  						\\
 24  & 4s$^2$4p$^4$($^1$D)4d   &  $^2$F$  _{7/2}$   &  3.80-02 & 2.688-03 & 2.798-03  &  5  - 24 M1 = 1.116+02, 8  - 24 M1 = 1.581+02				\\
 25  & 4s$^2$4p$^4$($^1$S)4d   &  $^2$D$  _{3/2}$   &  2.99-10 & 1.682-10 & 1.155-10  &  1  - 25 E1 = 3.373+09, 2  - 25 E1 = 2.572+09				\\
 26  & 4s$^2$4p$^4$($^1$S)4d   &  $^2$D$  _{5/2}$   &  5.34-09 & 2.211-09 & 2.472-10  &  1  - 26 E1 = 4.523+08  						\\
 27  & 4s$^2$4p$^4$($^1$D)4d   &  $^2$S$  _{1/2}$   &  3.55-12 & 4.120-12 & 4.808-12  &  1  - 27 E1 = 2.349+11  						\\
 28  & 4s$^2$4p$^4$($^3$P)4d   &  $^2$P$  _{3/2}$   &  4.41-12 & 4.849-12 & 5.472-12  &  1  - 28 E1 = 2.040+11  						\\
 29  & 4s$^2$4p$^4$($^3$P)4d   &  $^2$P$  _{1/2}$   &  2.65-10 & 4.884-12 & 4.777-12  &  2  - 29 E1 = 2.014+11  						\\
 30  & 4s$^2$4p$^4$($^3$P)4d   &  $^2$D$  _{5/2}$   &  3.17-12 & 3.725-12 & 4.209-12  &  1  - 30 E1 = 2.685+11  						\\
 31  & 4s$^2$4p$^4$($^3$P)4d   &  $^2$D$  _{3/2}$   &  3.25-12 & 3.823-12 & 4.240-12  &  2  - 31 E1 = 2.524+11  						\\
& & & & & & & & \\ \hline            										
\end{tabular}   								   					       
			      							   					       
\begin{flushleft}													       
{\small
GRASP2a: Singh {\em et al} \cite{sam} \\ 
GRASP2b: present calculations from the {\sc grasp} code with 470 levels\\
GRASP3: present calculations from the {\sc grasp} code with 3990 levels\\
															       
}															       
\end{flushleft} 
\end{document}